\PassOptionsToPackage{unicode}{hyperref}
\PassOptionsToPackage{hyphens}{url}
\PassOptionsToPackage{dvipsnames,svgnames,x11names}{xcolor}
\documentclass[
]{article}

\usepackage{amsmath,amssymb}
\usepackage{iftex}
\ifPDFTeX
  \usepackage[T1]{fontenc}
  \usepackage[utf8]{inputenc}
  \usepackage{textcomp} 
\else 
  \usepackage{unicode-math}
  \defaultfontfeatures{Scale=MatchLowercase}
  \defaultfontfeatures[\rmfamily]{Ligatures=TeX,Scale=1}
\fi
\usepackage{lmodern}
\ifPDFTeX\else  
\fi
\IfFileExists{upquote.sty}{\usepackage{upquote}}{}
\IfFileExists{microtype.sty}{
  \usepackage[]{microtype}
  \UseMicrotypeSet[protrusion]{basicmath} 
}{}
\makeatletter
\@ifundefined{KOMAClassName}{
  \IfFileExists{parskip.sty}{%
    \usepackage{parskip}
  }{
    \setlength{\parindent}{0pt}
    \setlength{\parskip}{6pt plus 2pt minus 1pt}}
}{
  \KOMAoptions{parskip=half}}
\makeatother
\usepackage{xcolor}
\ifx\paragraph\undefined\else
  \let\oldparagraph\paragraph
  \renewcommand{\paragraph}[1]{\oldparagraph{#1}\mbox{}}
\fi
\ifx\subparagraph\undefined\else
  \let\oldsubparagraph\subparagraph
  \renewcommand{\subparagraph}[1]{\oldsubparagraph{#1}\mbox{}}
\fi

\usepackage{longtable,booktabs,array}
\usepackage{calc} 
\usepackage{etoolbox}
\makeatletter
\patchcmd\longtable{\par}{\if@noskipsec\mbox{}\fi\par}{}{}
\makeatother
\IfFileExists{footnotehyper.sty}{\usepackage{footnotehyper}}{\usepackage{footnote}}
\makesavenoteenv{longtable}
\usepackage{graphicx}
\makeatletter
\def\maxwidth{\ifdim\Gin@nat@width>\linewidth\linewidth\else\Gin@nat@width\fi}
\def\maxheight{\ifdim\Gin@nat@height>\textheight\textheight\else\Gin@nat@height\fi}
\makeatother
\setkeys{Gin}{width=\maxwidth,height=\maxheight,keepaspectratio}
\makeatletter
\def\fps@figure{htbp}
\makeatother

\usepackage{tabularray}
\usepackage{arxiv}
\usepackage{orcidlink}
\usepackage{amsmath}
\usepackage[T1]{fontenc}
\usepackage{svg}
\usepackage{wrapfig}
\usepackage{soul}
\makeatletter
\@ifpackageloaded{caption}{}{\usepackage{caption}}
\AtBeginDocument{%
\ifdefined\contentsname
  \renewcommand*\contentsname{Table of contents}
\else
  \newcommand\contentsname{Table of contents}
\fi
\ifdefined\listfigurename
  \renewcommand*\listfigurename{List of Figures}
\else
  \newcommand\listfigurename{List of Figures}
\fi
\ifdefined\listtablename
  \renewcommand*\listtablename{List of Tables}
\else
  \newcommand\listtablename{List of Tables}
\fi
\ifdefined\figurename
  \renewcommand*\figurename{Figure}
\else
  \newcommand\figurename{Figure}
\fi
\ifdefined\tablename
  \renewcommand*\tablename{Table}
\else
  \newcommand\tablename{Table}
\fi
}
\@ifpackageloaded{float}{}{\usepackage{float}}
\floatstyle{ruled}
\@ifundefined{c@chapter}{\newfloat{codelisting}{h}{lop}}{\newfloat{codelisting}{h}{lop}[chapter]}
\floatname{codelisting}{Listing}

\makeatother
\makeatletter
\@ifpackageloaded{caption}{}{\usepackage{caption}}
\@ifpackageloaded{subcaption}{}{\usepackage{subcaption}}
\makeatother
\makeatletter
\@ifpackageloaded{tcolorbox}{}{\usepackage[skins,breakable]{tcolorbox}}
\makeatother
\makeatletter
\@ifundefined{shadecolor}{\definecolor{shadecolor}{rgb}{.97, .97, .97}}
\makeatother
\ifLuaTeX
\usepackage[bidi=basic]{babel}
\else
\usepackage[bidi=default]{babel}
\fi
\babelprovide[main,import]{american}

\def\languageshorthands#1{}
\ifLuaTeX
  \usepackage{selnolig}  
\fi
\usepackage[maxbibnames=99]{biblatex}
\IfFileExists{bookmark.sty}{\usepackage{bookmark}}{\usepackage{hyperref}}
\IfFileExists{xurl.sty}{\usepackage{xurl}}{} 
\hypersetup{
  pdftitle={Universal dimensions of visual representation},
  pdflang={en-US},
  colorlinks=true,
  linkcolor={blue},
  filecolor={Maroon},
  citecolor={Blue},
  urlcolor={Blue},
  pdfcreator={LaTeX via pandoc}}

\usepackage{lineno}
\title{Universal dimensions of visual representation}
\author{
    \textbf{Zirui Chen}~\orcidlink{0000-0003-3666-1719} \&
    \textbf{Michael F. Bonner}~\orcidlink{0000-0002-4992-674X}\\
    Department of Cognitive Science\\
    Johns Hopkins University, Baltimore,\ 21218\\
    \href{mailto:zchen160@jhu.edu}{zchen160@jhu.edu},
    \href{mailto:mfbonner@jhu.edu}{mfbonner@jhu.edu}
}
\begin{document}
\maketitle
\begin{abstract}

Do neural network models of vision learn brain-aligned representations because they share architectural constraints and task objectives with biological vision or because they learn universal features of natural image processing? We characterized the universality of hundreds of thousands of representational dimensions from visual neural networks with varied construction. We found that networks with varied architectures and task objectives learn to represent natural images using a shared set of latent dimensions, despite appearing highly distinct at a surface level. Next, by comparing these networks with human brain representations measured with fMRI, we found that the most brain-aligned representations in neural networks are those that are universal and independent of a network's specific characteristics. Remarkably, each network can be reduced to fewer than ten of its most universal dimensions with little impact on its representational similarity to the human brain. These results suggest that the underlying similarities between artificial and biological vision are primarily governed by a core set of universal image representations that are convergently learned by diverse systems.   

\end{abstract}
\ifdefined\Shaded\renewenvironment{Shaded}{\begin{tcolorbox}[breakable, sharp corners, frame hidden, interior hidden, enhanced, borderline west={3pt}{0pt}{shadecolor}, boxrule=0pt]}{\end{tcolorbox}}\fi

\section{Introduction}\label{introduction}

Deep neural networks have a remarkable ability to simulate the representations of biological vision \autocite{kriegeskorte2015deep, yamins2014performance, Conwell2022.03.28.485868, elmoznino2024high}. However, due to their immense complexity, the principles that govern the brain-aligned representations of deep networks remain poorly understood. 

A leading approach interprets neural network representations in terms of their architectures and task objectives, which are thought to function as key constraints on a network’s learned representations \autocite{yamins2016using, richards2019deep, cao2024explanatory, doerig2023neuroconnectionist, kanwisher2023using}. However, an alternative possibility is that the brain-aligned representations of neural networks are not contingent on specific optimization constraints but instead reflect universal aspects of natural image representation that emerge in diverse systems \autocite{guth2024on, huh2024platonic, elmoznino2024high}. 

Here we sought to determine if the representations that neural networks share with human vision are universal across networks. We examined over 200,000 dimensions of natural image representation in deep neural networks with varied designs. Our analyses revealed the existence of universal dimensions that are shared across networks and emerge under highly varied optimization conditions. Universal dimensions were observed across the full depth of network layers and across a variety of architectures and task objectives. Visualizations of these dimensions show that they do not simply encode low-level image statistics but also higher-level semantic properties. We next compared these dimensions to the representations of the human brain measured with fMRI, and we found that universal dimensions are highly brain-aligned and underlie conventional measures of representational similarity between neural networks and visual cortex. Together, these findings demonstrate the striking degree to which the shared properties of artificial and biological vision correspond to general-purpose representations that have little do to with the details of a network's architecture or task objective.

\section{Results}\label{results}

\subsection{Assessing universality and brain similarity}\label{}

We sought to compare two fundamental quantities of representational dimensions in neural networks: 1) their universality across varied networks and 2) their similarity to human brain representations. Here we briefly describe how we computed these two quantities. A more detailed description is provided in the Methods. 

\begin{figure}[H]
    \center
    \includegraphics{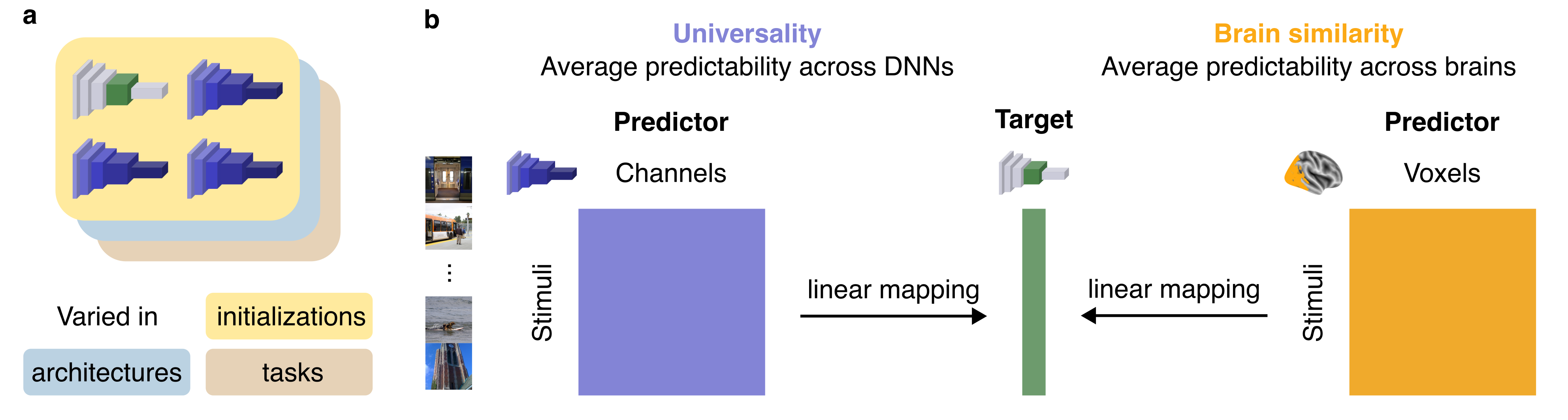}
    \caption{
        \textbf{Overview of method for computing universality and brain similarity of network dimensions.}
        \textbf{(a)} Four sets of deep neural networks were analyzed, including three sets of trained models that varied in either their random initializations, architectures, or task objectives and one set of untrained models with different initializations. 
        \textbf{(b)} Universality and brain similarity were defined as the average prediction accuracy of a latent dimension from a target network when using the activations of other networks or the fMRI activations of the human brain as predictors. Dimensions that can be consistently predicted from the representations of other networks have high universality. Dimensions that can be consistently predicted from the representations of the human brain have high brain similarity. 
    }
    \label{fig_1}
\end{figure}

\bigskip

As illustrated in Figure \ref{fig_1}, we characterized universality and brain similarity by examining the activations of networks and the human brain to a large and diverse set of natural images from the Microsoft Common Objects in Context (COCO) image database \autocite{lin2014microsoft}. For each latent dimension $d$ of a network’s activations (i.e., each principal component), we computed a universality metric by obtaining its average predictability from the activations of $m$ other networks:

\begin{equation}
\mathrm{Universality}_{d} = \operatorname{avg}(
\mathrm{r}_{d,1}, \
\mathrm{r}_{d,2}, \
\hdots, \
\mathrm{r}_{d,m}
)\,,
\end{equation}

where $\operatorname{avg}(\cdot)$ is the average and $r_{d,m}$ is the prediction accuracy of a cross-validated linear regression with dimension $d$ as the predictand and network $m$ as the predictor. We performed these analyses on the principal components (PCs) of a network’s activations so that each dimension $d$ is sampled from an orthogonal basis rather than from a set of potentially redundant neurons. The motivation for this metric is to quantify the average degree to which a dimension in one network is shared with the representations of other networks. Dimensions that are shared across all networks will have universality scores close to 1, while dimensions that only emerge under specific network configurations will have universality scores close to 0.

For each dimension $d$ of a network’s activations, we also computed a brain-similarity metric by obtaining its average predictability from the fMRI activations of $n$ human brains:

\begin{equation}
\mathrm{Brain \ similarity}_{d} = \operatorname{avg}(
\mathrm{r}_{d,1}, \
\mathrm{r}_{d,2}, \
\hdots, \
\mathrm{r}_{d,n}
)\,,
\end{equation}

where $\operatorname{avg}(\cdot)$ is the average and $r_{d,n}$ is the prediction accuracy of a cross-validated linear regression with dimension $d$ as the predictand and subject $n$ as the predictor. Here the motivation is to quantify the average degree to which a network dimension is shared with the representations of the human visual cortex. Dimensions that are shared between networks and humans will have brain similarity scores close to 1, while dimensions that are not shared with humans will have brain similarity scores close to 0. 

We examined the universality of representational dimensions among several sets of vision networks that varied in their random initializations, architectures, or task objectives. The specific networks used for these analyses are described in \ref{model-sets} and listed in Tables \ref{tab_2} and \ref{tab_3}. To assess brain similarity, we compared network representations with image-evoked fMRI responses from the Natural Scenes Dataset (NSD) \autocite{allen2022massive}, which is the largest existing fMRI dataset of natural scene perception in the human brain. We focused on a portion of this dataset that contains fMRI responses to 872 images shown to each of eight participants. This dataset is ideally suited for assessing whether the dimensions of natural image representations in neural networks can also be found in the representations of the human brain. For our main analyses, we focused on a general region of interest that included all voxels in visual cortex whose activity was modulated by the presentation of visual stimuli (Fig. \ref{fig_1}). 

\bigskip

\begin{figure}[H]
    \center
    \includegraphics{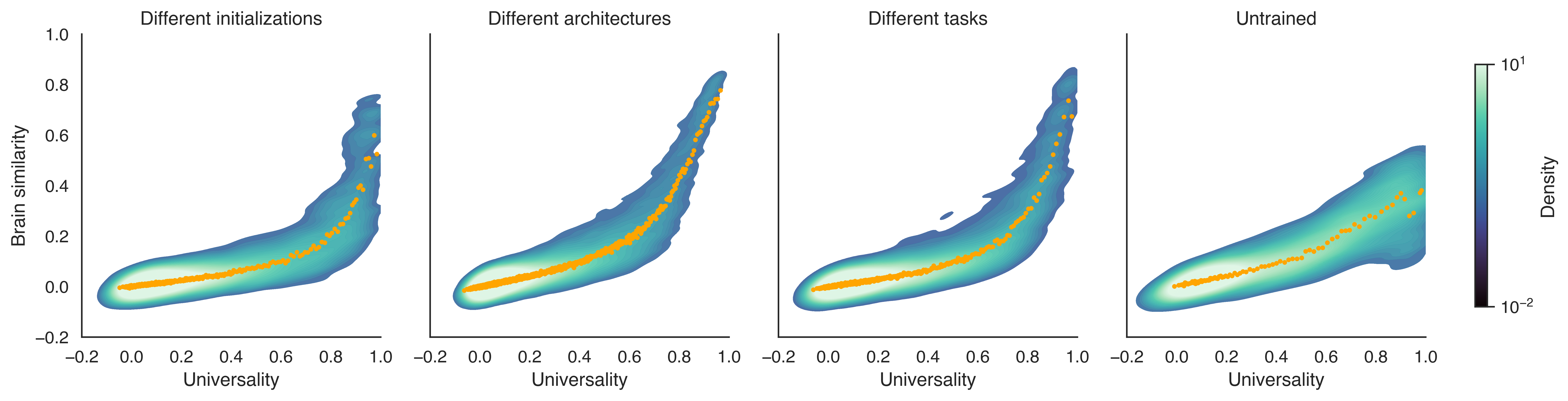}
    \caption{
        \textbf{Universality and brain similarity of network dimensions.}
        Universality and brain similarity were computed for representational dimensions in four sets of deep neural networks. These included three sets of trained networks with varied initializations, architectures, and task objectives and one set of untrained networks. These metrics were computed for the principal components of network activations extracted from the sampled layers of each network. Universality scores reflect the degree to which a representational dimension is shared across all networks in a set, and brain similarity scores reflect the degree to which a representational dimension is shared with the human visual system. Measurements of human visual cortex activity were obtained from the Natural Scenes fMRI Dataset using a general region of interest that included all visually responsive voxels \autocite{allen2022massive}. Universality and brain similarity scores are plotted for all analyzed network dimensions. These plots show the density of dimensions on a logarithmic scale, with densities computed using kernel density estimation. The orange dots show the mean universality and brain similarity scores for equally sized quantiles of 100 dimensions along the x-axis. These plots show similar trends for all three sets of trained models (the first three plots on the left). Specifically, they exhibit a high density of points near the origin, showing that most dimensions are idiosyncratic to each network and are not shared with the human brain. However, they also contain a subset of dimensions with exceptionally high universality and brain similarity scores. These latter dimensions correspond to representations that are consistently learned by all networks within a set and are also strongly shared with the visual representations of the human brain. In contrast, untrained networks (right panel) can also have shared dimensions, but these shared untrained dimensions have relatively weak brain similarity scores. 
    }
    \label{fig_2}
\end{figure}

\bigskip

\subsection{Universality across initialization weights}\label{}

We used universality and brain similarity to address a central question: Are there universal dimensions of natural image representation that are shared by neural networks and humans? We first performed these analyses for a setting in which we naturally expected to find shared network dimensions. Specifically, we examined universality among a set of networks that were initialized with different random weights but were otherwise identical (i.e., same architecture, task, and training data). These networks were 20 ResNet-18 architectures trained on image classification using the Tiny ImageNet dataset \autocite{he2016deep, schurholt2022model, le2015tiny}. For each dimension in each network layer, we computed its universality using the other networks as predictors. We examined network layers spanning the full range of model depth, except for the final classification layer. We iterated this analysis over a total of 36,596 dimensions across all networks. 

As shown in the left panel of Figure \ref{fig_2}, we observed universality scores spanning the full range from 0 to 1, with a high density of points around 0, indicating that most dimensions are idiosyncratic. The high density of idiosyncratic dimensions could reflect representations present at initialization that remain largely unchanged during training, or they could reflect unique representational strategies learned by specific network instances. In contrast, the universal dimensions at the other end of the scale reflect convergent representations that reliably emerge in all networks despite differences in their starting points. Notably, these universal dimensions account for a relatively small subset of the total number of network dimensions, as illustrated by the lower density of points at the high end of the universality axis.

We next compared these network dimensions to human brain representations, and we found that the universal dimensions exhibit exceptionally strong brain similarity scores (Fig. \ref{fig_2}). This demonstrates that among the many network dimensions examined here, it is only those that are invariably learned by networks with different initial conditions that are also strongly shared with the representations of the human visual system. 

In follow-up analyses, we found that the trend in Figure \ref{fig_2} was consistently observed in each network layer (Fig. \ref{fig_3}), each individual network (Fig. \ref{sup_1}), each individual fMRI subject (Fig. \ref{sup_2}), and in multiple regions of interest in visual cortex (Fig. \ref{sup_3}). The observation of this effect in all network layers shows that universality is not restricted to low-level features in early layers but instead extends across the full depth of the network. We emphasize that the universality and brain similarity metrics in our analysis pipeline are not guaranteed to be related to one another. Using simulated data, we can trivially obtain a range of universality and brain similarity scores while observing no positive relationship between these two metrics (Fig. \ref{sup_4}). Furthermore, we re-computed the universality and brain similarity metrics using other mapping procedures to ensure that our findings are not contingent on the specific regression method that we used. These analyses showed that the variation in universality and brain similarity across network dimensions is highly consistent when using unregularized regression instead of ridge regression and even when using a one-to-one mapping procedure without any regression-based re-weighting (\ref{alternative}; Fig. \ref{sup_11}). Thus, the observed trends are highly robust and are not contingent on regularization or regression-based mapping in general. Together, these findings show that among a preponderance of idiosyncratic network dimensions, there exists a smaller subset of highly convergent dimensions that are learned by different network instances and are shared with human vision.

\subsection{Universality across architectures and tasks}\label{}

We next sought to determine if universal dimensions can be detected among networks with varied architectures and tasks. To this end, we quantified universality and brain similarity for two sets of models. The first was a set of models with different architectures but trained on the same task. Specifically, we examined 19 networks trained to perform ImageNet classification using varied architectures, including convolutional models, vision transformers, and MLP-Mixers. The second was a set of 9 models with the same architecture (ResNet-50) but trained on different tasks. The tasks included object classification and a variety of self-supervised tasks, such as contrastive learning, identifying image rotations, and solving jigsaw puzzles. For all models, we examined layers spanning the full network depth, except for the final layer. Further details about the models can be found in Tables \ref{tab_2} and \ref{tab_3}. In total, we examined 149,743 dimensions in the set of varied architectures and 43,132 dimensions in the set of varied tasks. 

The findings for these two sets of models were surprisingly consistent with those observed for models with varied initializations (Fig. \ref{fig_2}). We again found that most dimensions are idiosyncratic (i.e., specific to a model) and not shared with the brain, as shown by the high density of points near the origin. Again, we also found that a subset of dimensions exhibit exceptionally high scores on both the universality and brain similarity metrics. These latter dimensions correspond to representations that reliably emerge across many models despite variations in their architectures and the tasks that they were trained to perform. Furthermore, the generality of these representations extends beyond artificial vision, as they are also strongly shared with the representations of the human visual system. Remarkably, these findings are highly similar when considering networks that vary in either architectures or task objectives. This suggests that the underlying similarities among the representations of these networks---as well as their similarities to human vision---are only weakly influenced by architecture and task but instead reflect highly general properties of image representations in deep networks. 

\bigskip

\begin{figure}[H]
    \center
    \includegraphics{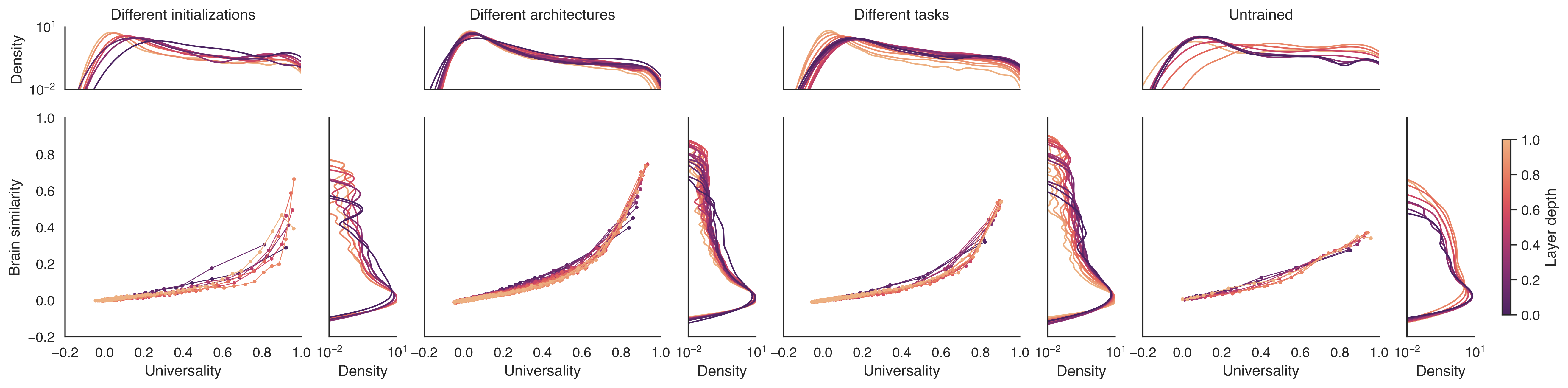}
    \caption{
    \textbf{Universality and brain similarity across network layers.}
    These plots show the universality and brain similarity scores for individual network layers spanning the full depth of each network. Four sets of deep neural networks were examined, including three sets of trained networks with varied initializations, architectures, and tasks and one set of untrained networks. The analyses are the same as in Figure \ref{fig_2}, but here the results are plotted as the average values for individual layers, which are labeled according to their relative depth. Further details of these network layers are included in the supplement file \href{https://github.com/zche377/universal_dimensions/blob/main/src/lib/models/model_layers.csv}{model\_layer.csv}. Average universality and brain similarity scores were computed for equally sized quantiles of 100 dimensions along the x-axis for each layer. Panels on the sides of each plot show the density of dimensions on a logarithmic scale computed using kernel density estimation. As in Figure \ref{fig_2}, these plots exhibit a high density of points near the origin, which means that across all sampled layers, most dimensions are idiosyncratic and are not shared with the human brain. However, the three sets of trained networks (first three plots on the left) also contain a subset of dimensions at the right end of each plot that have exceptionally high universality and brain similarity scores. Importantly, these layer-wise plots show that a consistent trend is observed across all sampled layers and that universal dimensions are not restricted to early network layers. 
    }
    \label{fig_3}
\end{figure}

We also note that there is a striking paucity of points in the upper left quadrant of the plots in Figure \ref{fig_2}. Points in this quadrant would correspond to representations shared with the brain but learned only by networks with specific optimization constraints---namely, specific architectures or tasks. The lack of points in this quadrant suggests that the details of architectures and tasks have a relatively minor role in shaping the brain-aligned representations of neural networks. 

In follow-up analyses, we again found that these results are highly robust---they were observed in each network layer (Fig. \ref{fig_3}), each individual network (Fig. \ref{sup_1}), each individual fMRI subject (Fig. \ref{sup_2}), and in multiple regions of interest in visual cortex (Fig. \ref{sup_3}). Together, these findings reveal the remarkable degree to which vision systems with varied architectures and task objectives can nonetheless converge on a set of general-purpose representations that are shared not only across models but also between artificial and biological vision. 

\subsection{Universality across untrained networks}\label{}

Our analyses thus far have focused on sets of trained neural networks, and we have interpreted the dimensions in the upper right quadrant of the plots in Figure \ref{fig_2} as \emph{learned} representations. However, we expect that shared dimensions can also be found among sets of untrained models due to statistical regularities in the activations that natural images elicit in networks with random filters. We thus wondered whether our findings for the trained networks could be explained by the statistics of image activations alone---without any need for learning---or whether they diverge from the trends observed in randomly initialized networks. To address this question, we examined 20 ResNet-18 architectures that were randomly initialized with the same seeds as the trained models presented in the left panel of Figure \ref{fig_2}. We followed the same procedures as in the preceding analyses. For each dimension in each network layer, we computed its universality using the other networks as predictors, and we iterated this analysis over a total of 9,413 dimensions from all networks. 

These analyses showed that, as expected, there is a wide range of universality scores for the untrained network dimensions, with some dimensions that are found in all networks (Fig. \ref{fig_2}). These universal dimensions of untrained networks correspond to representations that consistently emerge when propagating natural images through a hierarchy of random convolutional filters. They are thus due to image statistics alone and not learned representational properties. However, importantly, the relationship between universality and brain similarity diverges from the relationship that was observed for trained models. Specifically, for untrained networks, we observe a shallow and approximately linear relationship between universality and brain similarity, whereas for trained networks, brain similarity exhibits a sharp nonlinear increase at the high end of the universality axis. As a result, the shared dimensions of untrained networks have substantially lower brain similarity scores than the shared dimensions of trained networks. As in the previous sets of analyses, we again found that these results were consistent in each network layer (Fig. \ref{fig_3}), each individual network (Fig. \ref{sup_1}), each individual fMRI subject (Fig. \ref{sup_2}), and in multiple regions of interest in visual cortex (Fig. \ref{sup_3}). In sum, when comparing the trends for trained and untrained networks, the findings demonstrate that the universal dimensions of trained networks reflect \emph{learned} representational properties that cannot be explained by image statistics and random features alone.

Finally, we performed an analysis to determine if the observed trends in brain similarity might be affected by the reliability of the fMRI data. Specifically, we wondered if low brain similarity scores might reflect low reliability in the fMRI data rather than a mismatch between a network and the brain. The brain similarity scores in our analyses can only be as high as the reliability of the underlying fMRI data across subjects. Hence, it is possible to compute a straightforward metric of between-subject reliability and adjust the brain similarity scores by this metric. As shown in Fig. \ref{sup_12}, the observed trends in brain similarity for all four sets of models are qualitatively unchanged after adjusting for reliability.

\subsection{Universality and the visual hierarchy}\label{}

Previous work has shown that in many neural networks trained on natural images, the first layer contains general-purpose V1-like filters tuned to orientation, frequency, and color, whereas subsequent layers contain filters that appear to be increasingly specialized \autocite{NIPS2012_c399862d, yosinski2015understanding}. This suggests the possibility that universality may only be prominent in early network layers and then rapidly diminish across the network hierarchy. To address this possibility, we examined the universality and brain similarity of network representations in individual layers along the full depth of each network. Figure \ref{fig_3} shows the results of these analyses for all sets of models. Across all sets of trained models, we did not find a strong layer effect but instead, relatively similar distributions of universality scores at all sampled layers, with highly universal dimensions detected even in the deepest layers that we examined. Furthermore, these analyses show that the relationship between universality and brain similarity is consistent across layers. Thus, these findings suggest that at all levels of network depth, we can find general-purpose representations that are reliably learned by diverse networks and are strongly shared with the human brain. 

\begin{figure}[H]
    \center
    \includegraphics{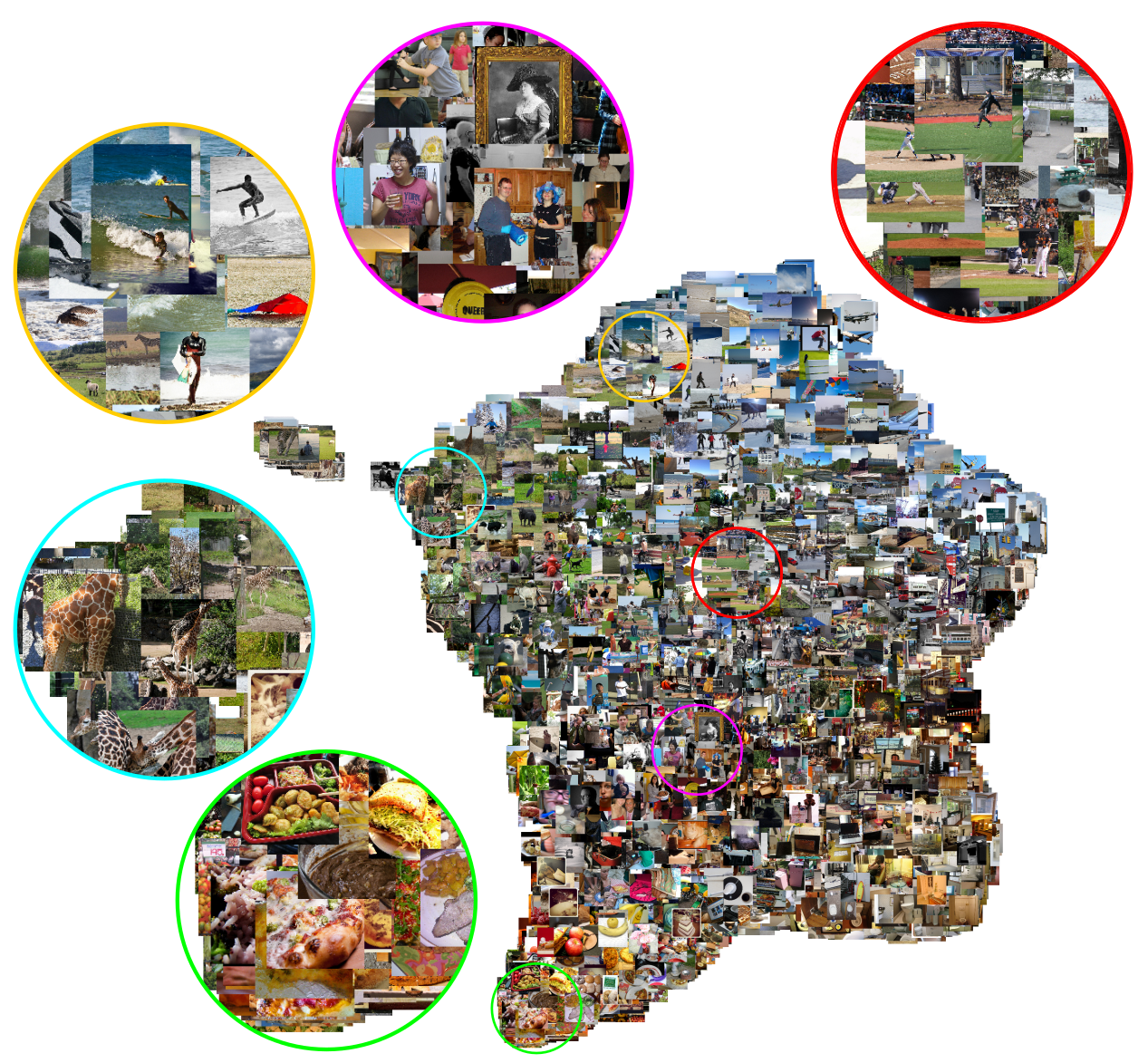}
    \caption{
    \textbf{Two-dimensional visualization of high-level universal representations.}
    Image activations for the 100 most universal dimensions from a high-level network layer were embedded in two dimensions using uniform manifold approximation and projection. Specifically, image activations were obtained for the top 100 dimensions with the highest universality scores in the penultimate layer from the set of models trained on different tasks. For visualization purposes, this figure only includes images shown to a single subject. This plot shows that universal dimensions do not simply reflect low-level image features but instead capture high-level properties that group images into semantically related clusters, some of which are highlighted here, including animals, food, sports, and people. In contrast, Figure \ref{sup_5} shows a visualization of the 100 \emph{least} universal dimensions from the same network layer, and it shows no clear semantic organization. 
    }
    \label{fig_4}
\end{figure}

\bigskip

\subsection{Universal dimensions and high-level image properties}\label{}

The findings from the previous section show that universal dimensions are not restricted to early network layers. However, these findings do not directly address the question of whether the shared dimensions of later layers represent high-level semantic properties or low-level image statistics, such as luminance gradients and spatial frequency distributions. To address this question, we performed exploratory visualization analyses of the universal dimensions in later network layers. Specifically, we examined the penultimate layer of the models from the varied-task set, as shown in the middle right panel of Figure \ref{fig_2}. We focused on the varied-task set because the models in this set have the same ResNet-50 architecture, which allowed us to examine representations from the same targeted layer in all networks. We concatenated the dimensions from the penultimate layer across all networks and ranked their universality scores. We then selected the top 100 dimensions with the highest universality scores and visualized the image representations of these 100 dimensions in a 2D space using uniform manifold approximation and projection (UMAP) \autocite{mcinnes2018umap-software}. 

As shown in Figure \ref{fig_4}, these representations exhibit rich high-level organization, with images grouped into clusters of semantically related items, such as people, sports, animals, and food. To verify our interpretation of this visualization, we used representational similarity analysis (RSA) to compare the UMAP embeddings of all ~73K images from the NSD dataset with semantic embeddings based on image captions. The semantic embeddings were obtained from a sentence transformer and averaged across all five human-annotated captions \autocite{song2020mpnet, reimers-2019-sentence-bert}, and representational dissimilarity matrices were computed using Euclidean distance. This analysis revealed a substantial RSA correlation of $r=0.35$ between the semantic embeddings and the UMAP embeddings of the universal dimensions. For comparison, we also generated UMAP embeddings of the 100 dimensions with the lowest universality scores from the same ResNet-50 layer. We found that these idiosyncratic dimensions exhibit no clear semantic organization (Fig. \ref{sup_5}) and no correlation with the semantic embeddings ($r=0.00$). We also generated UMAP embeddings of the top 100 universal dimensions in the penultimate layer of the untrained-network set, which is shown in the rightmost panel of Figure \ref{fig_2}. In contrast to the trained networks, the universal dimensions of untrained networks emphasize prominent low-level features, such as coarse luminance gradients (Fig. \ref{sup_6}), which again are not correlated with the semantic embeddings ($r=0.03$). Together, these visualization analyses show that the universal representations of high-level layers in trained networks encode high-level image properties that group images into semantically meaningful clusters. This suggests that there are common organizing principles of high-level image semantics that are universally learned. 

\subsection{Universal dimensions and representational similarity analysis}\label{}

\begin{figure}[H]
    \center
    \includegraphics{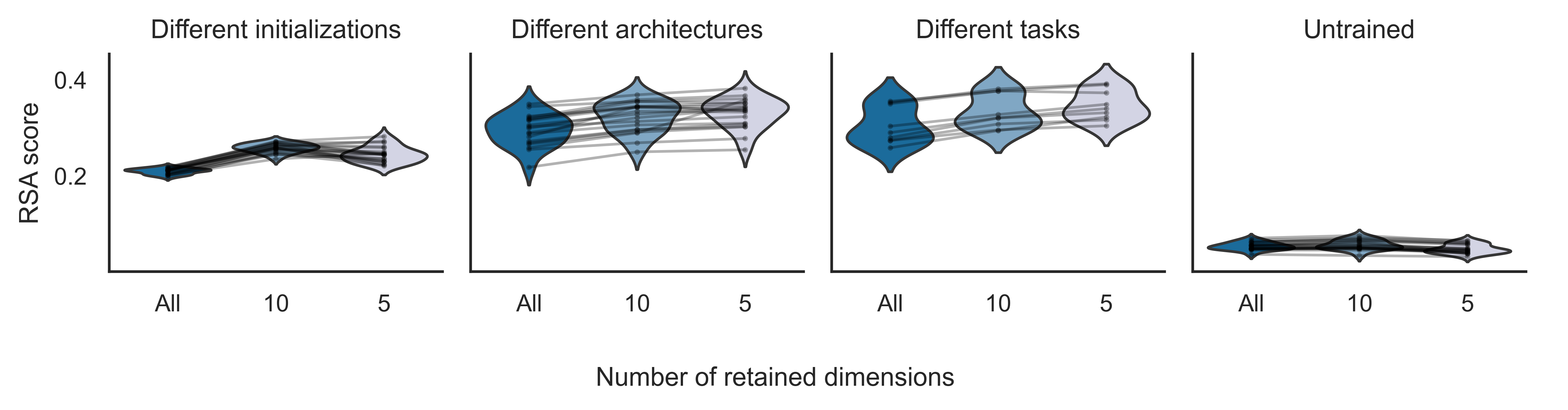}
    \caption{
    \textbf{Universal dimensions underlie the results of conventional representational similarity analyses}
    Representational similarity analysis (RSA) was used to compare the representations of neural networks and visual cortex. These analyses were performed using the same general region of interest in visual cortex and the same sets of neural networks as in Figures \ref{fig_2} and \ref{fig_3}. Representational dissimilarity matrices (RDMs) were created by calculating Pearson correlation distances for pairwise comparisons of image representations within each network and each fMRI subject. RSA scores were obtained by calculating the Spearman correlation between the RDMs for a network and an fMRI subject. These RSA scores were averaged across subjects. For each network, the best-performing layer was selected using a set of training data, and a final RSA score was computed on held-out test data. These analyses show the results of RSA for networks whose representations were either intact or reduced to subspaces of their top ten or five universal dimensions. In these plots, each dot is a network, and lines connect different versions of a network containing either all, ten, or five dimensions. The violin plots show distributions of RSA scores across networks. Even after drastically reducing the networks to just ten or five universal dimensions, the RSA scores exhibit little or no decrease---in fact, for all three sets of trained networks, the RSA scores slightly improve. These results demonstrate that conventional measures of representational similarity between neural networks and visual cortex are largely driven by the subspaces of universal dimensions contained within each network. Similar trends were observed within each individual subject and in other regions of interest, as shown in Figures \ref{sup_7} and \ref{sup_8}.
    }
    \label{fig_5}
\end{figure}

Our findings thus far show that the universal dimensions of networks can be strongly predicted from human brain representations. We next sought to evaluate the effect of universal dimensions on a conventional RSA. Specifically, we performed a targeted analysis to determine if universal dimensions drive the representational similarity scores obtained from comparisons of neural networks with visual cortex. To do so, we conducted a standard RSA on networks with representations reduced to low-dimensional subspaces of their most universal dimensions. We performed this analysis for all sets of networks examined in Figure \ref{fig_2}. Following the RSA procedures in \autocite{Conwell2022.03.28.485868}, we split the stimuli into training and test sets. We selected the best-performing layer from each network on the training set and computed the final RSA score for the selected layers on the test set. We then reduced each network to a subset of its most universal dimensions and computed RSA scores for these reduced model representations. We analyzed the same general region of interest in visual cortex as in the preceding analyses. We found that even when the networks are reduced to just ten or five universal dimensions, their RSA scores exhibit little or no decrease---in fact, for all three sets of trained networks, they slightly improve (Fig. \ref{fig_5}). Similar results were observed when performing these analyses in individual subjects (Fig. \ref{sup_7}) and in other regions of interest (Fig. \ref{sup_8}). These findings suggest that the representational similarities between neural networks and visual cortex are largely driven by subspaces of network dimensions that are universal. 

\section{Discussion}\label{discussion}

Our work reveals universal dimensions of natural image representation that are learned by artificial vision systems and are shared with the human brain. These dimensions emerge in diverse neural networks despite variation in their architectures and task objectives. The role of these dimensions in vision appears to be general-purpose---they are not specialized for any single task but instead support many downstream objectives. Universal representations are found at all levels of visual processing in deep networks, from low-level image properties in early layers to high-level semantics in late layers. Together, these findings suggest that machines and humans share a set of general-purpose visual representations that can be learned under highly varied conditions. 

Deep learning is now the standard framework for computational modeling in neuroscience, and many previous efforts have sought to understand these deep learning models in terms of their specialization: that is, what objectives they are specialized for, and what specific network characteristics underlie their similarities to the brain \autocite{yamins2016using, richards2019deep, cao2024explanatory, doerig2023neuroconnectionist, kanwisher2023using}. Our work views the representations of deep networks from a different perspective. Rather than searching for specific model characteristics that might be associated with stronger alignment with the brain, we sought to discover the elements of network representations that are instead \emph{invariant} across models. Using this approach, we found that crucial aspects of deep network representations---those that are most strongly shared with the human brain---are, to a remarkable degree, independent of the network characteristics that many previous studies have emphasized. The invariance of these representations implies that they are not primarily governed by the details of a network’s architecture or task objective but instead by more general principles of natural image representation in deep vision systems \autocite{guth2024on, huh2024platonic}. 

\bigskip

In our analyses, we focused on the principal components (PCs) of each network’s representations. We chose to focus on PCs for two reasons. First, this approach allowed us to specifically characterize the proportion of distinct (i.e., orthogonal) dimensions that have high universality, which we found to be remarkably small relative to the hundreds of thousands of dimensions we analyzed. Second, many previous studies comparing representations across networks have focused on principal axes of variance, and both empirical and theoretical work suggests that the information a network learns is primarily reflected in its lower-rank PCs \autocite{elmoznino2024high, kornblith2019similarity, guth2023rainbow}. However, there are alternative dimensions that could be explored, including non-orthogonal bases, sparse bases, or nonlinear manifold coordinates \autocite{klindt2023identifying, cunningham2023sparse, NEURIPS2019_cfcce062}. An intriguing question for future work is whether alternative coordinate systems could reveal an even greater degree of universality across networks. 

It is important to note that universality and PC rank are fundamentally distinct attributes of representational dimensions, and one should not assume that all low-rank PCs are universal. For example, variations in tasks, architectures, and training procedures can dramatically alter the shape of the representational eigenspectrum and, thus, the interpretation of PC ranks \autocite{elmoznino2024high, prince2024manipulating, garrido2023rankme}, and it appears that some modern architectures and self-supervised learning objectives can yield low-rank PCs that are highly idiosyncratic \autocite{pmlr-v243-robinson24a}. To illustrate this in our own data, we have plotted the distribution of universality scores for three different decades of PC ranks (Fig. \ref{sup_14}). As expected, we find that low-rank PCs typically have high universality scores, but, importantly, within each decade of PC ranks, the universality scores span a wide range. In fact, universality scores for the top ten PCs can range as low as those in the third decade of PC ranks. This means that while universal dimensions tend to be low-rank, not all low-rank PCs are universal. 

The universal dimensions that we examined in this work emerge in models that are trained with a shared visual diet, despite differences in architecture and task objective. An important direction for future work is to understand whether models trained with different sets of natural images also learn shared representations. More broadly, it may be possible to quantify how the overlap of training data distributions governs the degree to which networks learned shared representations \autocite{flovik2024quantifying}. As a first step in this direction, we performed a preliminary analysis of the shared dimensions across a set of publicly available networks matched on architecture and task objective but trained on distinct sets of natural images (including object-centric images, complex scenes with multiple objects, faces, and large-scale places) \autocite{konkle2022self}. The findings suggest that even when networks are trained on highly distinct sets of natural images, they nonetheless learn a subset of strongly shared representational dimensions, and as in our other analyses, it is these shared dimensions that are most similar the representations of the human visual system (Fig. \ref{sup_13}). This preliminary analysis suggests that the universal dimensions examined here reflect highly general properties of natural images, but more work is needed to precisely determine the relationship between universality and the statistics of training set distributions. 

Our findings suggest several exciting directions for future work. First, our approach could be extended beyond vision models to examine the representational dimensions that are shared across vision and language. Previous work has shown that language-model embeddings of object names and scene captions are predictive of image representations in high-level visual cortex \autocite{carlson2014emergence, bonner2021object, doerig2024visualrepresentationshumanbrain}. An open question is whether networks trained on language data alone learn the same universal dimensions of natural scene representation as image-trained networks. Second, our findings show that universal dimensions emerge in networks despite differences in the tasks that the networks are optimized to perform. This raises the intriguing possibility that universal dimensions could be hard-coded into networks at initialization, potentially making the learning process faster and more data efficient. Third, while previous work has revealed similarities between the visual cortex representations of humans and monkeys \autocite{kriegeskorte2008matching}, we still know little about the degree to which representational dimensions may be universal or species-specific across mammalian vision. This question could be addressed by applying our approach to recordings of cortical responses to the same stimuli in different species. 

In sum, our results show that the most brain-aligned representations of visual neural networks are universal and independent of a network’s specific characteristics. What fundamental principles might explain the convergence of networks to universal dimensions? Theories of efficient coding suggest that frequency- and orientation-tuned filters are consistently observed in the first layer of vision systems because they constitute efficient bases that are adapted to the statistics of natural images \autocite{olshausen1996emergence, simoncelli2001natural}. It remains an open question whether this efficient-coding hypothesis can be extended to a deep hierarchy, which could potentially explain universal dimensions as a consequence of optimal image encoding. An alternative possibility is that deep networks learn shared representations of the true generative factors in the visual world---e.g., the objects, materials, contexts, and optical phenomena that make a scene. This could be the case if the optimal strategy for solving challenging tasks on natural stimuli is to learn the invariant properties of reality \autocite{huh2024platonic}, and it would suggest that the universal dimensions detected here reflect a shared internal model of the visual environment in machines and humans.

\clearpage

\section{Methods}\label{methods}

\subsection{Natural Scenes Dataset}\label{nsd}

\subsubsection{Stimuli and experimental design}\label{visual-stimuli}

The Natural Scenes Dataset (NSD) is a large-scale publicly available fMRI dataset on human vision that is described in detail in a previous report \autocite{allen2022massive}. Here we briefly review the key attributes of this dataset. The NSD study sourced color natural scene stimuli from the Microsoft Common Objects in Context (COCO) database \autocite{lin2014microsoft} and collected 7T fMRI responses (1.8mm voxels, 1.6s TR) from eight adult subjects who viewed these stimuli while performing a continuous recognition memory task, namely to respond if the presented image was seen before in the experiment. Each subject viewed approximately 10,000 stimuli with three repetitions, though some subjects saw fewer stimuli because they did not complete all scanning sessions. Among the stimuli, 872 "shared" images were viewed by all subjects at least once. We used these shared images and the corresponding fMRI data for our main analyses. 

\subsubsection{Data preprocessing}\label{data-preprocessing}

We used the NSD single-trial betas, preprocessed in 1.8-mm volume space and denoised with the GLMdenoise technique (\texttt{version 3}; \texttt{betas\_fithrf\_GLMdenoiseRR}). Subsequently, the betas were transformed to z-scores within each individual scanning session, as recommended by the authors \autocite{allen2022massive}. For all analyses, we used the averaged betas across repetitions.

\subsubsection{Regions of interest}\label{regions-of-interest}

Our main analyses focused on the \texttt{nsdgeneral} region of interest (ROI), which includes a large swath of visual cortex. This ROI, as defined in the NSD study \autocite{allen2022massive}, contains all voxels that were reliably modulated by the presentation of visual stimuli, comprising approximately 15,000 voxels in each subject. We also conducted follow-up analyses in smaller ROIs, including the \texttt{ventral}, \texttt{parietal}, \texttt{lateral} streams, using the "streams" ROIs provided in the NSD dataset. 

\subsection{Deep neural networks}\label{dnn}

\subsubsection{Model sets}\label{model-sets}

We examined four sets of DNN models in the main results to probe the universality of representational features across different factors of variation:
\begin{itemize}
  \item Random seeds in trained models
  \item Architectures
  \item Task objectives 
  \item Random seeds in untrained models.
\end{itemize}

The first set included 20 pretrained ResNet-18 models examined in a previous study of model hyperparameters \autocite{schurholt2022model}. These models were initialized with unique random seeds and trained on Tiny ImageNet \autocite{le2015tiny}. We extracted features from nine rectified linear unit (ReLU) layers that span the depth of the ResNet-18 architecture, except for the final output layer. In total, we extracted 36,596 dimensions from this set.

The second set included 19 pretrained models with varied architectures, which included various convolutional networks, transformers, and MLP-Mixers (Table \ref{tab_2}). All were trained on ImageNet for object classification \autocite{russakovsky2015imagenet} and obtained from the torchvision library \autocite{paszke2019pytorch,rw2019timm}. Given the variety of architectures in this set, the sampled layers included ReLU, normalization, attention, multi-layer perceptron, and other model-specific operations. The sampled layers span the full range of layer depth in each model (additional details are provided in Table \ref{tab_2}). The complete list of these layers are included in the supplement file \href{https://github.com/zche377/universal_dimensions/blob/main/src/lib/models/model_layers.csv}{model\_layer.csv}. A total of 149,743 dimensions were extracted from this set.

The third set included 9 pretrained ResNet-50 models from the torchvision library \autocite{paszke2019pytorch} and the VISSL model zoo \autocite{goyal2021vissl}. These models were trained to perform a variety of tasks on ImageNet images \autocite{russakovsky2015imagenet} (Table \ref{tab_3}). A total of 43,132 dimensions were extracted from all ReLU layers, except for the final output layer.

The fourth set included 20 untrained ResNet-18 models \autocite{he2016deep} with different random weights, which were created using Kaiming normal initialization \autocite{he2015delving}. Each model had a unique random seed. A total of 9,413 dimensions were extracted from all ReLU layers. Note that the lower number of dimensions here relative to the set of trained models with different random seeds is due to the low-rank activation matrices of untrained networks. 

We also performed a supplementary analysis in a fifth set that included 4 instance-prototype contrastive learning (IPCL) models trained with varied visual diets \autocite{konkle2022self} (Table \ref{tab_4}). These models used a modified AlexNet architecture \autocite{krizhevsky2012imagenet} with group-initialization instead of batch normalization layers. A total of 2,758 dimensions were extracted from all ReLU layers, except for the final output layer.

\subsubsection{Feature extraction}\label{feature-extraction}

Before computing our universality and brain-similarity metrics, we first needed to extract a set of feature activations from each model layer. We sought to quantify these metrics for the \emph{features} representations in each network rather than for the spatial representations. We thus applied global max-pooling to remove spatial information from the activations of each model layer. For convolutional networks, pooling was applied across the height and width dimensions, and for networks with patch embeddings, pooling was applied across patch dimensions. To extract all orthogonal dimensions from each model layer, we first performed PCA on the activations to the 72,128 "unshared" images from NSD, retaining all PCs up to the matrix rank, computed with the default procedure in \texttt{torch.linalg.matrix\char`_rank} in PyTorch \autocite{paszke2019pytorch}. We then transformed the 872 "shared" images from NSD to the PC basis and computed universality and brain similarity metrics for each PC. 

\subsection{Metrics}\label{metrics}

\subsubsection{Universality}\label{universality}

Our universality metric estimates the degree to which a representational dimension is shared across multiple DNNs. For a given dimension in a \emph{target} network, we used cross-validated ridge regression to predict its activations as a linear combination of the activations from another \emph{predictor} network. We performed this analysis using the same "shared" NSD images that were used to compute brain similarity (described in \ref{brain-similarity}). The regressors consisted of activations concatenated across all sampled layers of the predictor network. The procedure for cross-validated ridge regression is described in \ref{ridgecv}. We computed the mean Pearson correlation between the predicted and actual responses of the target dimension across all cross-validation folds, and we repeated this process, using every network other than the target as the predictor. We then obtained the universality score by taking the median correlation across all predictor networks. We used median instead of mean to ensure that the final summary statistic was not driven by a small subset of predictor networks with exceptionally high or low scores. This entire procedure was repeated to obtain universality scores for all dimension in all sampled layers of the target network, and it was then repeated with each network as the target. 

\subsubsection{Brain similarity}\label{brain-similarity}

Our brain similarity metric estimates the degree to which a dimension in a DNN can be predicted from fMRI responses measured in human visual cortex. Given a target dimension from a network, we used cross-validated ridge regression to predict its activations as a linear combination of the trial-averaged fMRI responses to the "shared" images in a single subject. We computed the mean Pearson correlation between the predicted and actual responses of the target dimension across all cross-validation folds, and we repeated this procedure for all fMRI subjects. Brain similarity is defined as the mean score across all subjects.

\subsubsection{Cross-validated ridge regression}\label{ridgecv}

We computed universality and brain similarity scores using ridge regression with a nested cross-validation design. The outer loop of this cross-validation design had five folds. We fit the parameters of the ridge regression on four folds of training data. We first selected the optimal ridge penalty for each target dimension from zero and values with equal logarithmic spacing between \(10^{-3}\) to \(10^{4}\). The optimal ridge penalty was the one that yielded that best performance when applying leave-one-out cross-validation to the training data. We then fit the regression weights using the full set of training data and the optimal ridge penalty, and we applied these regression weights to generate predicted responses in the held-out fold of test data. Performance was evaluated as the correlation between the predicted and actual responses on the held-out test data. This procedure was repeated using all five folds as held-out test data, and the performance scores were average across folds.

\subsubsection{Alternative mapping methods}\label{alternative}

We compared two other mapping methods with the ridge regression used for the main results (\ref{ridgecv}), both conducted with the same five-fold cross-validation design. The first method was ordinary least square regression without any regularization. The second method is one-to-one mapping: in the training folds, we identified the column in the predictors with highest Pearson correlation with the target feature, and we then computed the Pearson correlation between the selected predictor column and target feature in the test fold.

\subsubsection{Representational similarity analysis}\label{rsa}

We computed conventional representational similarity analysis (RSA) scores for comparisons of networks and visual cortex \autocite{kriegeskorte2008representational}, adapting the procedure described in \autocite{Conwell2022.03.28.485868}. We split the 872 "shared" images into a training set and a test set of 436 images each. In each set, representational dissimilarity matrices (RDMs) were created by calculating Pearson correlation distances for pairwise comparisons of image representations within each network layer and each fMRI subject. For each network layer, we computed RDMs using the same globally pooled channel activations that were used to compute universality and brain similarity scores. RSA scores were obtained by calculating the Spearman correlation between the RDMs for a network layer and an fMRI subject, and these scores were averaged across subjects. For each network, the best-performing layer was selected based on the RSA scores in the training set, and a final RSA score was computed for each network using the selected layer in the held-out test set. We next examined the contribution of universal dimensions to the RSA scores by reducing each network to the subspace spanned by its top ten or five universal dimensions. Specifically, we reconstructed the test-set activations of each network using only the top ten or five most universal dimensions, and we re-computed the final RSA score on these reconstructed test data. 

%

%

\section{Data availability}\label{data}

The Natural Scenes Dataset is available at \url{https://naturalscenesdataset.org/} \autocite{allen2022massive}.

\section{Code availability}\label{code}

Code for all analyses in this study is available at \url{https://github.com/zche377/universal_dimensions}.

\section{Acknowledgements}\label{acknowledgements}

This research was supported in part by a JHU Catalyst Award to MFB and grant NSF PHY-2309135 to the Kavli Institute for Theoretical Physics (KITP).

\section{Declaration of interests}\label{declaration}
The authors declare no competing interests.

\clearpage

\printbibliography

\clearpage

\section{Supplementary material}\label{supplementary-material}

\newcommand{\beginsupplement}{%
    \setcounter{table}{0}
    \renewcommand{\thetable}{S\arabic{table}}%
    \setcounter{figure}{0}
    \renewcommand{\thefigure}{S\arabic{figure}}%
 }
 \beginsupplement

\begin{figure}[H]
    \center
    \includegraphics{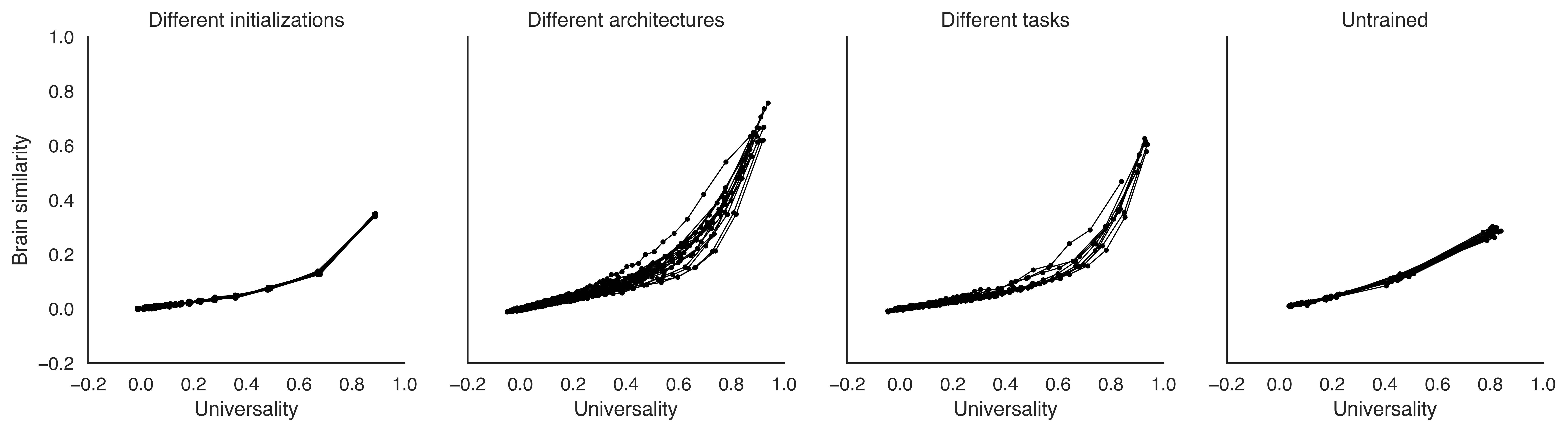}
    \caption{
    \textbf{Universality and brain similarity for individual networks.}
    These plots show the universality and brain similarity scores for individual networks. Four sets of deep neural networks were examined, including three sets of trained networks with varied initializations, architectures, and tasks and one set of untrained networks. The analyses are the same as in Figure \ref{fig_2}, but here the results are plotted as the average values for individual networks. Average universality and brain similarity scores were computed for equally sized quantiles of 100 dimensions along the x-axis for each network. As in Figure \ref{fig_2}, these plots show that a highly consistent trend is observed across all networks and that universal dimensions are not restricted to a subset of networks.
    }
    \label{sup_1}
\end{figure}
\clearpage

\begin{figure}[H]
    \center
    \includegraphics[width=0.65\textwidth]{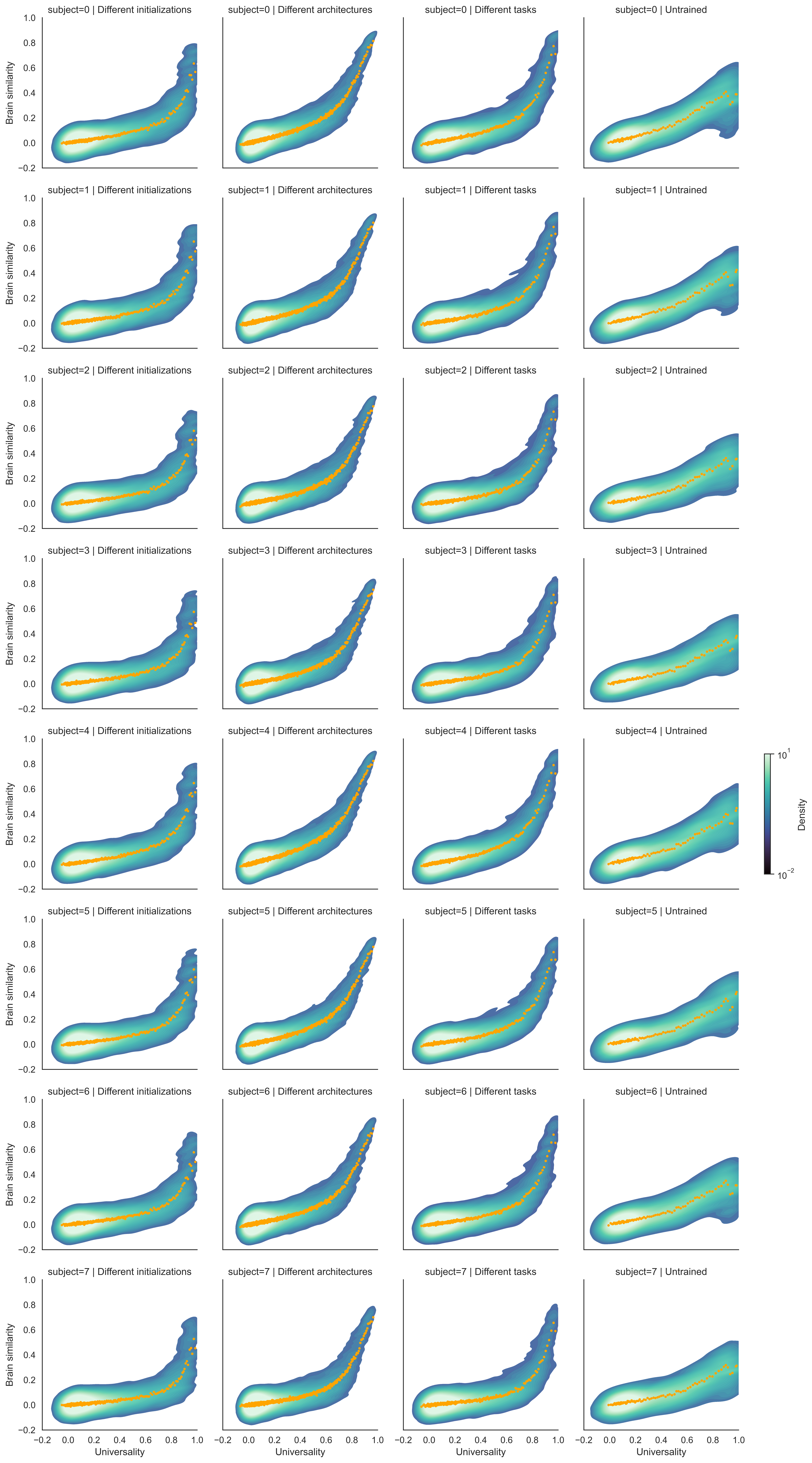}
    \caption{
    \textbf{Universality and brain similarity of network dimensions in individual subjects.}
    These plots show the relationship between universality and brain similarity in each subject from the fMRI dataset. The analyses are the same in Figure \ref{fig_2} but without averaging the brain similarity scores across subjects. As in Figure \ref{fig_2}, these plots show the density of dimensions on a logarithmic scale computed using kernel density estimation. The orange dots show the mean universality and brain similarity scores for equally sized quantiles of 100 dimensions along the x-axis. These results demonstrate that the relationship between universality and brain similarity is highly consistent and robustly detected in all individual subjects. 
    }
    \label{sup_2}
\end{figure}
\clearpage

\begin{figure}[H]
    \center
    \includegraphics{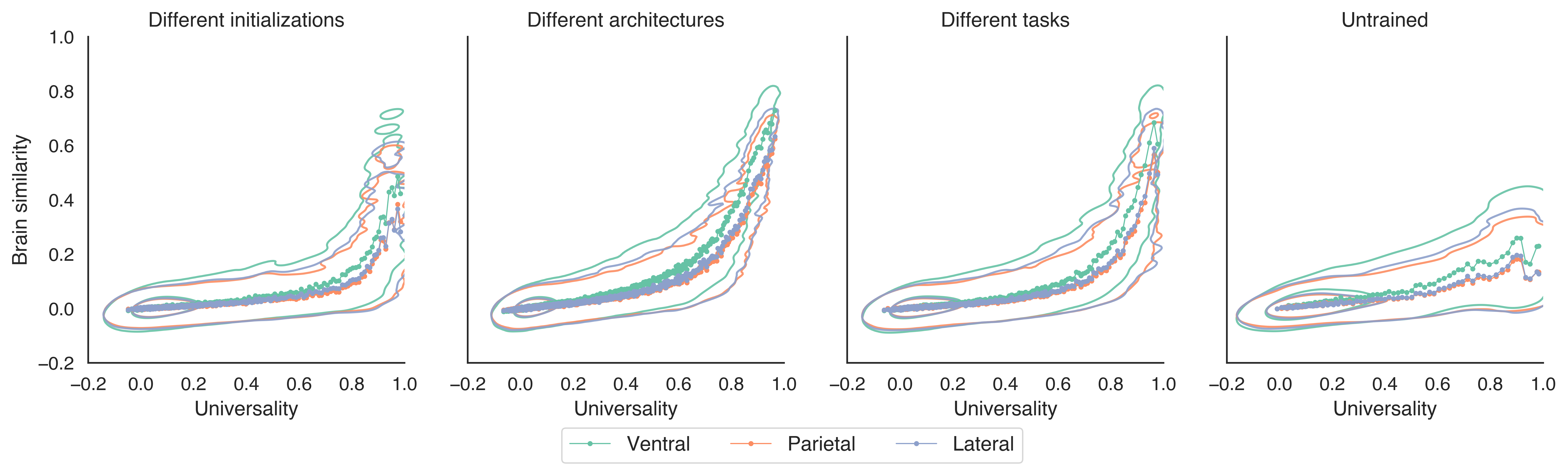}
    \caption{
    \textbf{Universality and brain similarity for multiple regions of interest in visual cortex.}
    These plots show the universality and brain similarity scores for three regions of interest: the ventral, parietal, and lateral streams. These regions are based on the “streams” masks as defined by the authors of the Natural Scenes Dataset study \autocite{allen2022massive}. The analysis is the same as in Figure \ref{fig_2}, but here brain similarity scores are computed using the three stream regions rather than the large \texttt{nsdgeneral} region from our main analyses. To visualize the results for all three regions in a single plot, the contours of the kernel density estimate plots are displayed here, rather than their density values. Average universality and brain similarity scores were computed for equally sized quantiles of 100 dimensions along the x-axis for each region. These plots show that a highly consistent trend is observed across all regions. 
    }
    \label{sup_3}
\end{figure}
\clearpage


\begin{figure}[H]
    \center
    \includegraphics{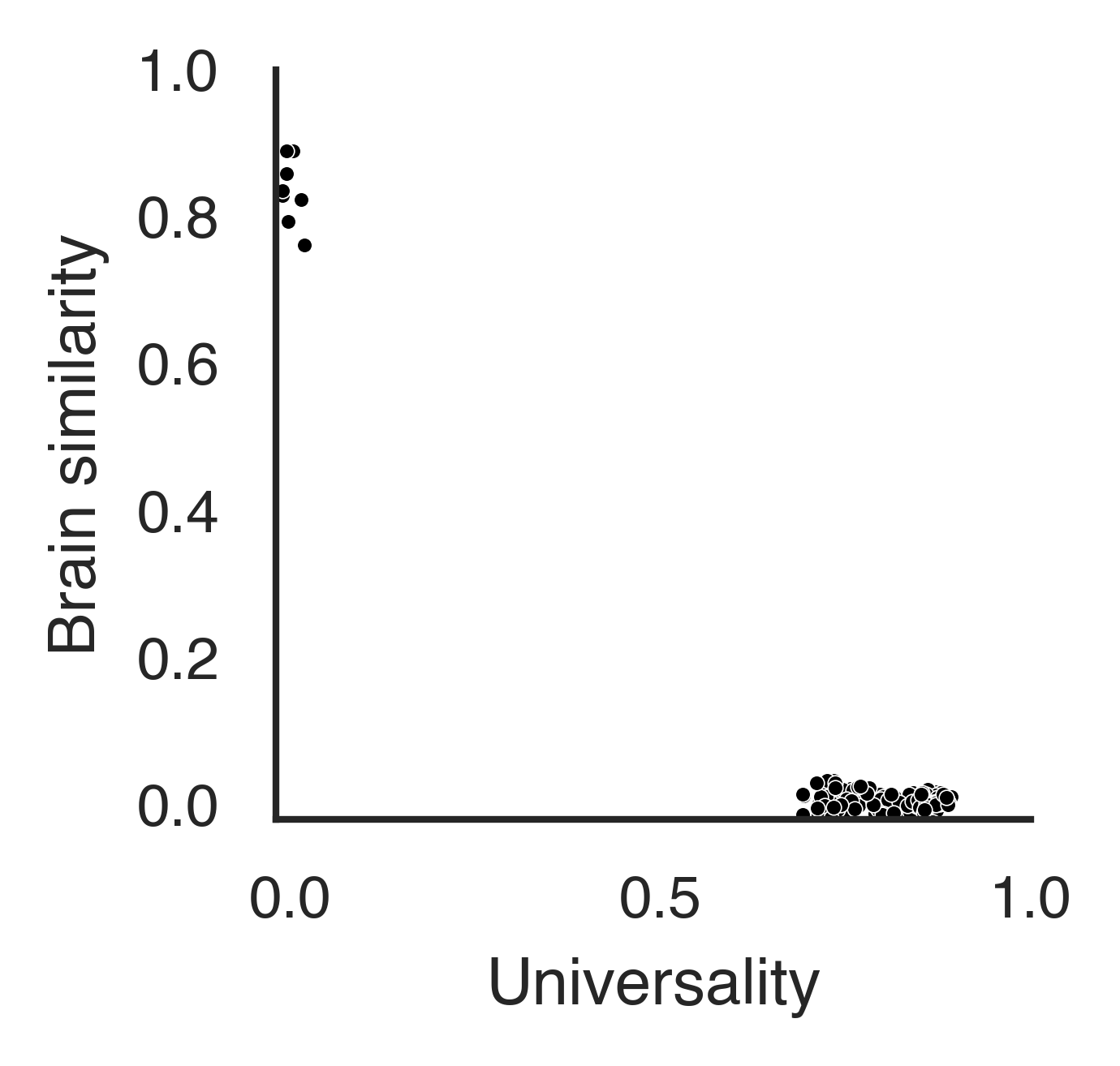}
    \caption{
    \textbf{Universality and brain similarity are not intrinsically correlated.}
    Universality and brain similarity scores were computed on simulated data to demonstrate that these metrics are not intrinsically correlated and can be trivially dissociated from one another. Data were generated for 20 simulated “subjects” and “networks.” All subjects and a single network were created from a common matrix of orthonormal random variables with added Gaussian noise. The remaining networks were created from another matrix of orthonormal random variables with added Gaussian noise. These simulated data yield a subset of dimensions with low universality and high brain similarity and another set of dimensions with high universality and low brain similarity. The plot limits for both axes are set to [-0.02, 1] to make the data points near 0 visible. 
    }
    \label{sup_4}
\end{figure}
\clearpage

\begin{figure}[H]
    \center
    \includegraphics[width=0.75\textwidth]{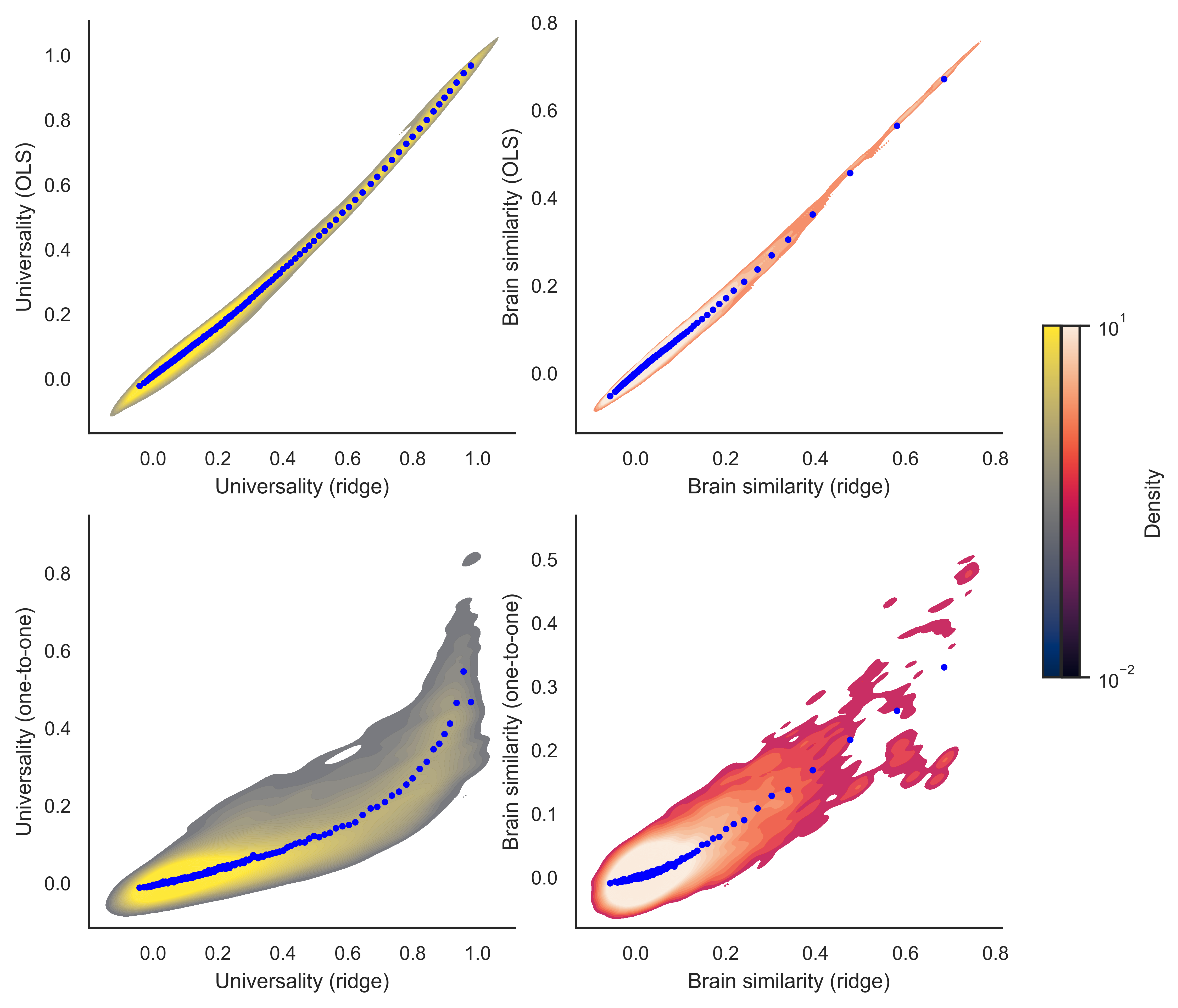}
    \caption{
    \textbf{Universality and brain similarity computed with different mapping methods}
    Universality and brain similarity were computed for representational dimensions in the set of 20 ResNet-18 architectures trained on image classification using the Tiny ImageNet dataset \autocite{he2016deep, schurholt2022model, le2015tiny}, initialized with varied seeds. Plots in each row compared the metrics computed with the default ridge regression method to those computed with ordinary least square (OLS) regression or one-to-one mapping (\ref{alternative}). These plots show the default metrics on the x-axis and the alternative metrics on the y-axis, with the density of dimensions computed using kernel density estimation. The blue dots show the mean universality and brain similarity scores for equally sized quantiles of 200 dimensions along the x-axis. The results show that these metrics are not strongly contingent on the use of regularized regression and that similar trends are observed even without regression-based reweighting (i.e., with one-to-one mapping). Note that, as expected, the use of one-to-one mapping makes all the values lower, but the resulting values are nonetheless strongly correlated with those obtained with the ridge regression procedure.
    }
    \label{sup_11}
\end{figure}
\clearpage

\begin{figure}[H]
    \center
    \includegraphics{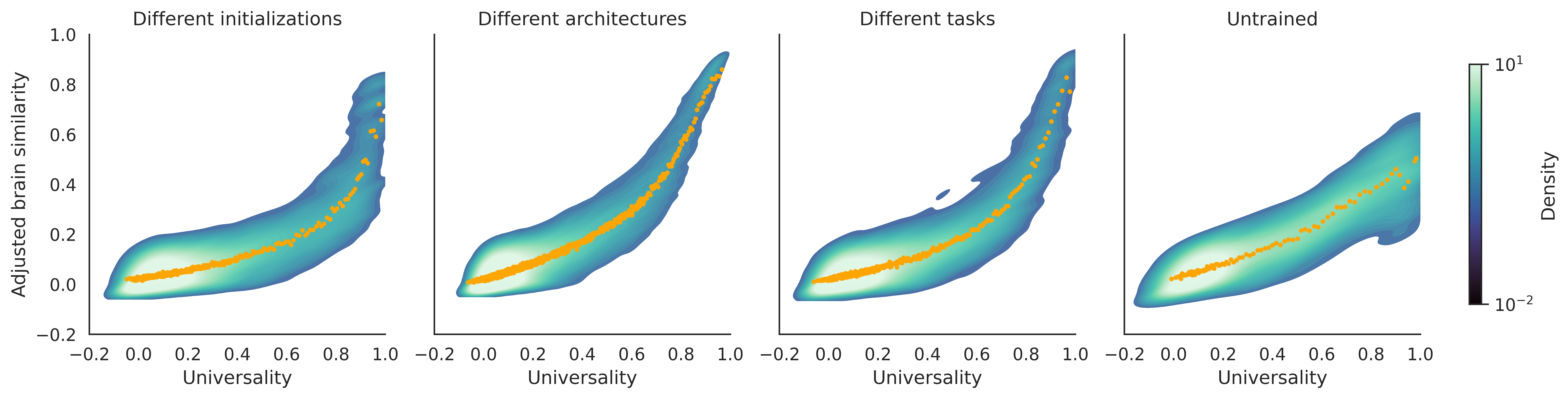}
    \caption{
    \textbf{Universality and adjusted brain similarity of network dimensions}
    These plots show the relationship between universality and brain similarity adjusted by the inter-subject reliability of the fMRI data. Because brain similarity is calculated as a correlation between a network dimension and a linear projection of the fMRI data, we can compute the relevant fMRI reliability as the average between-subject correlation of the projected fMRI responses. These between-subject correlations were calculated on each test fold and averaged across folds. Noise-ceiling adjusted brain similarity scores were computed by dividing the original brain similarity by the square root of the between-subject reliability (with negative values set to zero). Four sets of deep neural networks were examined, including three sets of trained networks with varied initializations, architectures, and tasks and one set of untrained networks. As in Figure \ref{fig_2}, these plots show the density of dimensions on a logarithmic scale computed using kernel density estimation. The orange dots show the mean universality and adjusted brain similarity scores for equally sized quantiles of 100 dimensions along the x-axis. The results are highly similar to those shown in Figure \ref{fig_2} using unadjusted brain similarity scores.
    }
    \label{sup_12}
\end{figure}
\clearpage

\begin{figure}[H]
    \center
    \includegraphics{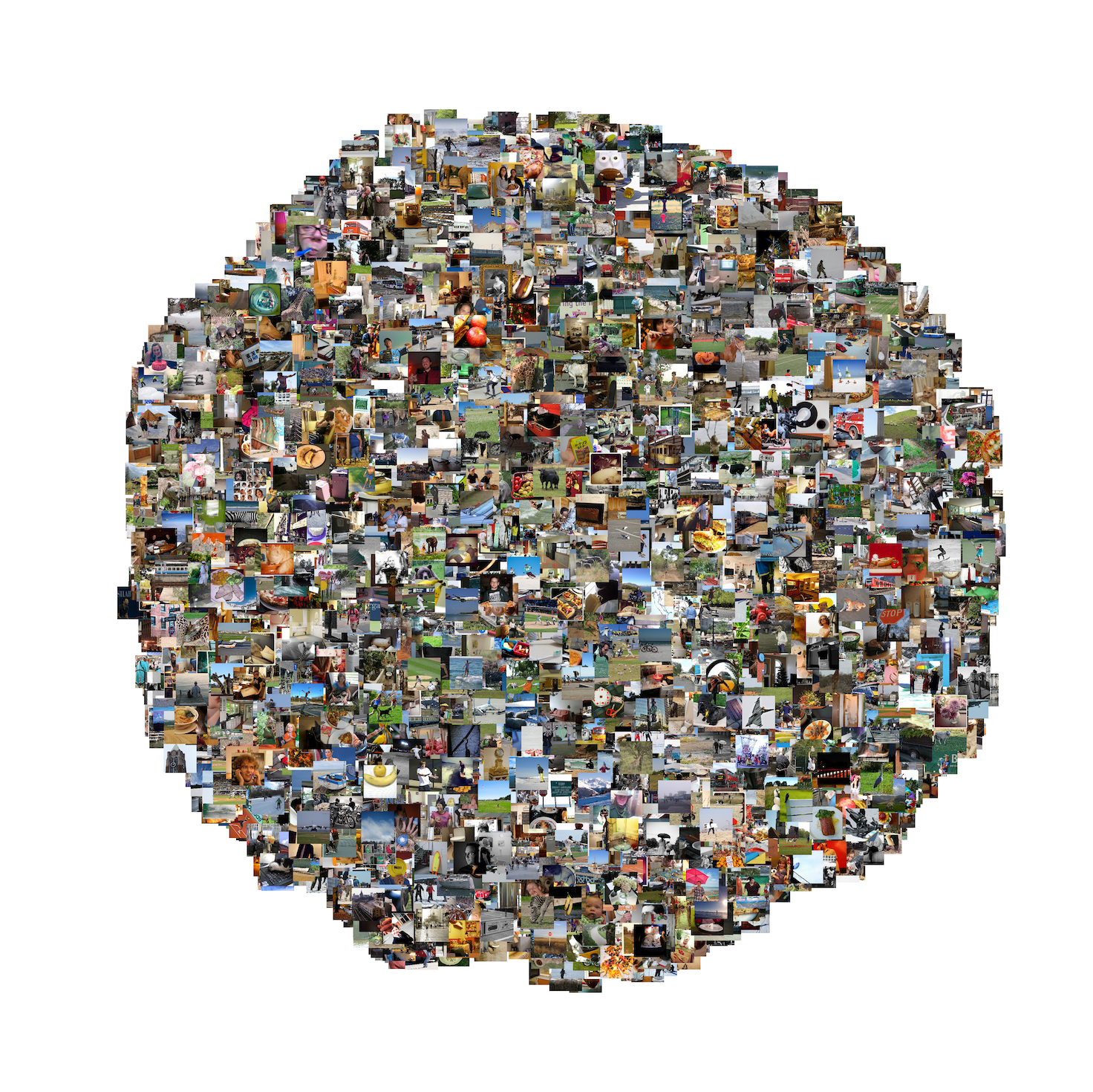}
    \caption{
    \textbf{Two-dimensional visualization of model-specific representations.}
    Image activations for the 100 \emph{least} universal dimensions from a high-level network layer were embedded in two dimensions using uniform manifold approximation and projection. Specifically, image activations were obtained for the 100 dimensions with the lowest universality scores in the penultimate layer from the set of ResNet-50 models trained on different tasks. In contrast to the universal dimensions visualized in Figure \ref{fig_4}, this plot shows no clear semantic organization.
    }
    \label{sup_5}
\end{figure}
\clearpage

\begin{figure}[H]
    \center
    \includegraphics{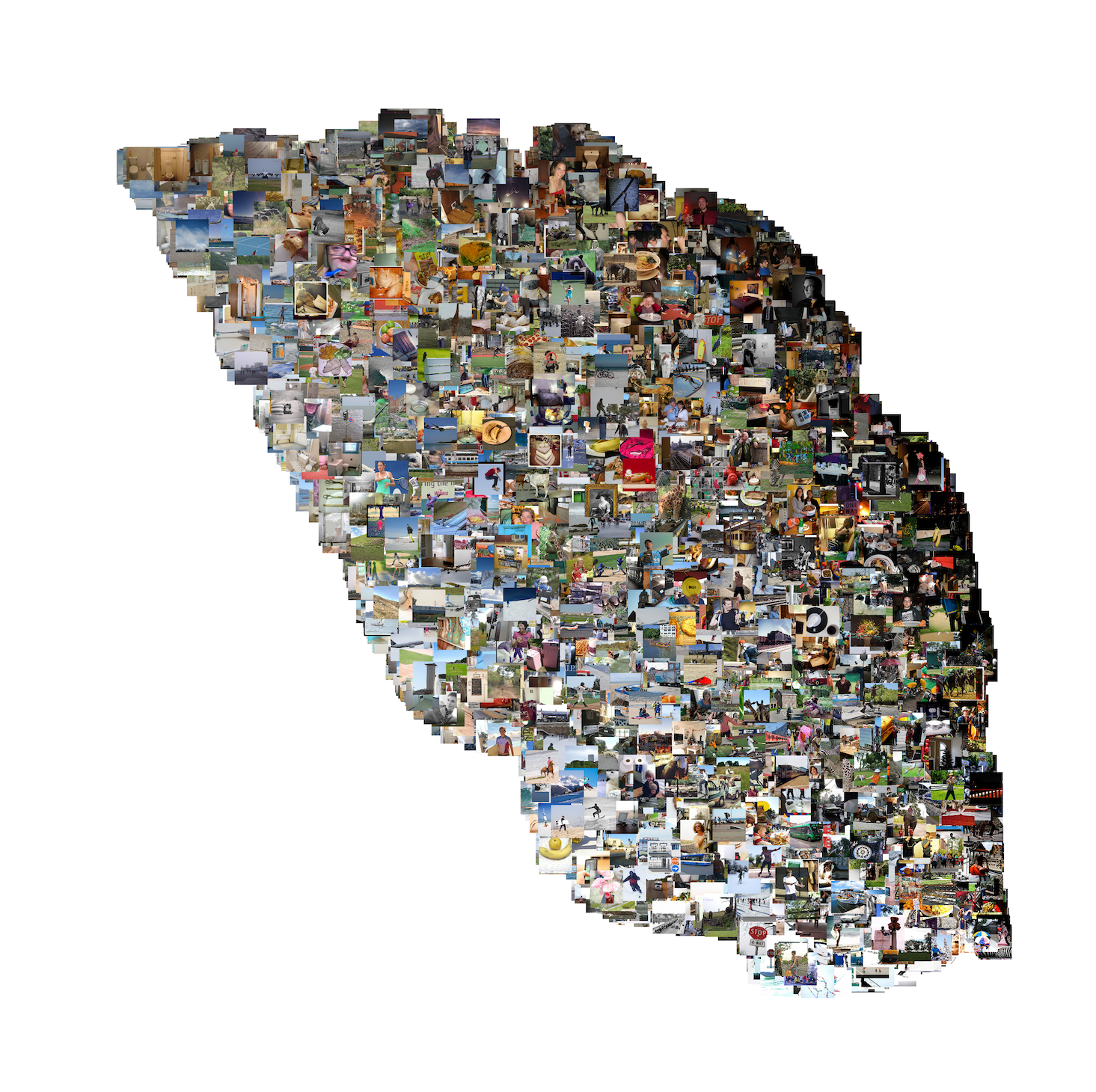}
    \caption{
    \textbf{Two-dimensional visualization of untrained models.}
    Image activations for the 100 most universal dimensions of untrained networks were embedded in two dimensions using uniform manifold approximation and projection. Specifically, image activations were obtained for the top 100 dimensions with the highest universality scores in the penultimate layer from the set of untrained ResNet-18 models with different random weights. In contrast to the universal dimensions visualized in Figure \ref{fig_4}, this plot shows no clear semantic organization. Instead, the universal dimensions of untrained networks appear to emphasize low-level image properties, as demonstrated by the strong luminance gradient from left to right in this plot.
    }
    \label{sup_6}
\end{figure}
\clearpage

\begin{figure}[H]
    \center
    \includegraphics[width=0.75\textwidth]{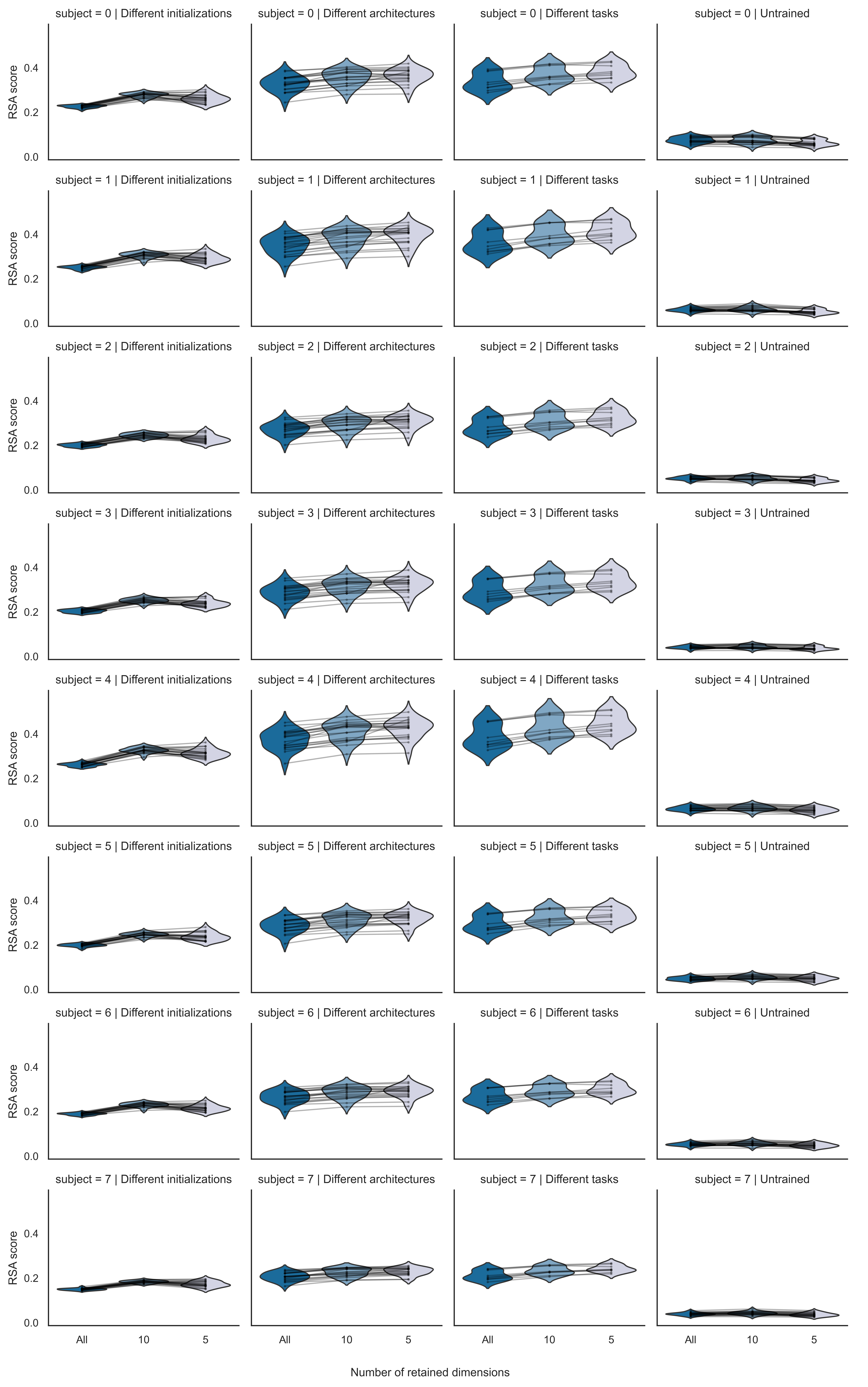}
    \caption{
    \textbf{Universal dimensions underlie the results of representational similarity analyses for individual subjects.}
    These plots show the results of representational similarity analyses (RSA) comparing networks with each subject from the fMRI dataset. The analyses are the same as in Figure \ref{fig_5} but without averaging the RSA scores across subjects. As in Figure \ref{fig_5}, each dot is a network, whose representations were either intact or reduced to subspaces of their top ten or five universal dimensions, and the violin plots show distributions of RSA scores across networks. These results demonstrate that the subspaces of universal dimensions within each network consistently drive the representational similarity between neural networks and visual cortex across all individual subjects.
    }
    \label{sup_7}
\end{figure}
\clearpage

\begin{figure}[H]
    \center
    \includegraphics{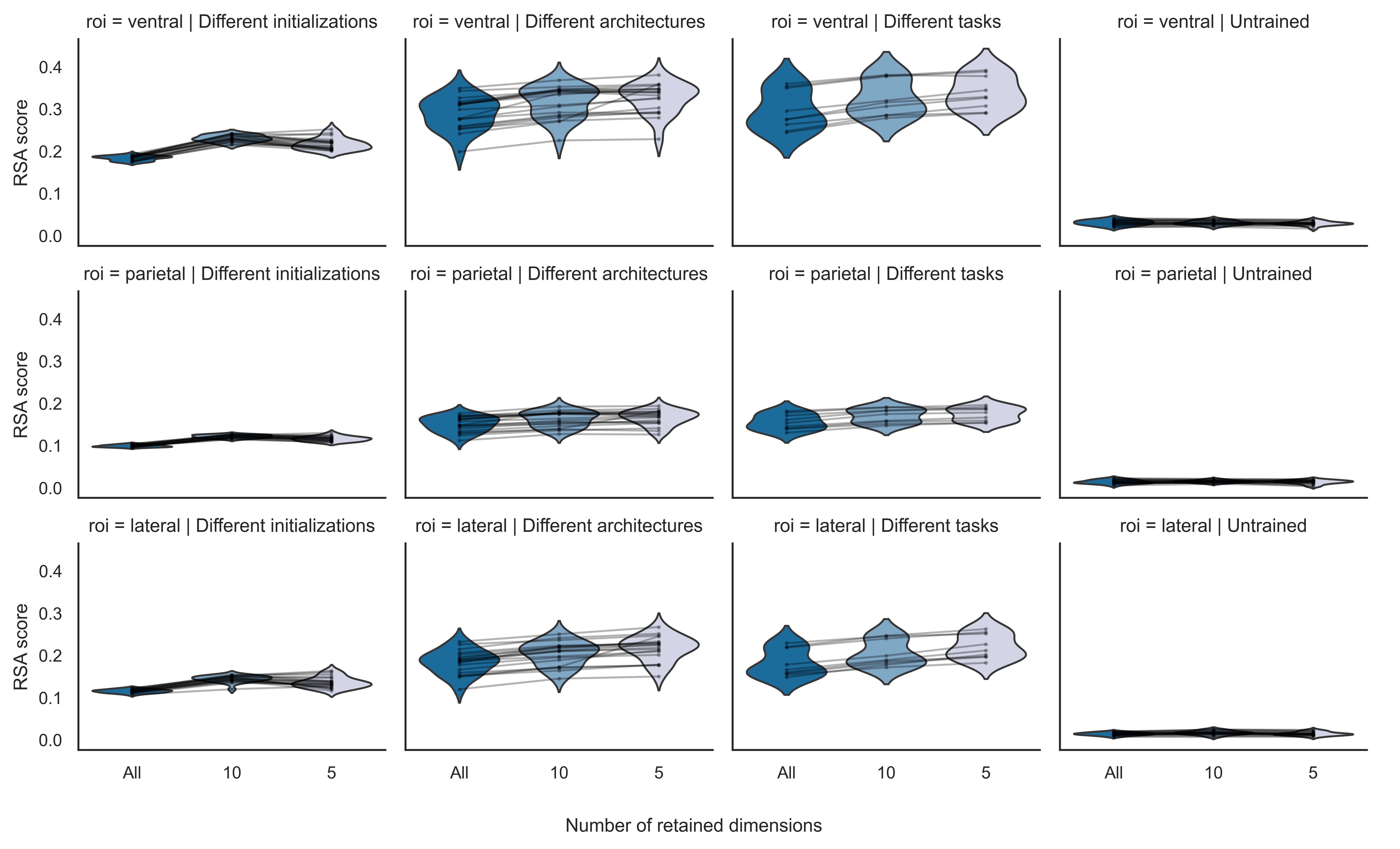}
    \caption{
    \textbf{Universal dimensions underlie the results of representational similarity analyses in multiple regions of interest.}
    These plots show the results of representational similarity analyses (RSA) comparing networks with the fMRI responses from three regions of interest: the ventral, parietal , and lateral streams. These regions are based on the “streams” masks as defined by the authors of the Natural Scenes Dataset study \autocite{allen2022massive}. The analyses are the same as in Figure \ref{fig_5}, but here the representational dissimilarity matrices (RDM) for the fMRI data are computed using the three stream regions rather than the large nsdgeneral region from our main analyses. As in Figure \ref{fig_5}, each dot is a network, whose representations were either intact or reduced to subspaces of their top ten or five universal dimensions, and the violin plots show distributions of RSA scores across networks. These results demonstrate that the subspaces of universal dimensions within each network consistently drive the representational similarity between neural networks and visual cortex across all regions.
    }
    \label{sup_8}
\end{figure}
\clearpage

\begin{figure}[H]
    \center
    \includegraphics{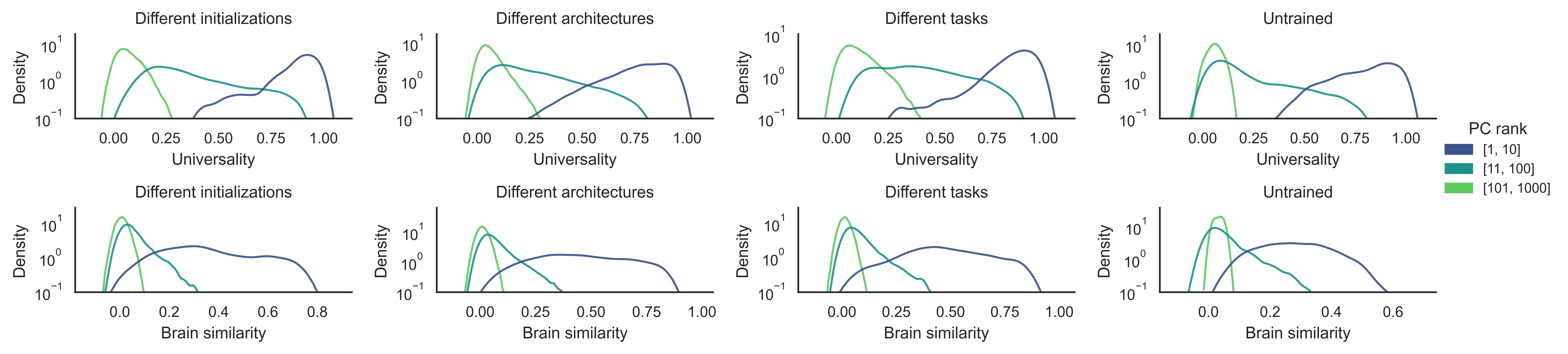}
    \caption{
    \textbf{Distributions of universality and brain similarity for three decades of PC ranks}
    These plots show distributions of universality and brain similarity scores for PCs extracted from each model layer in the four sets of networks shown in Figure \ref{fig_2}. These distributions are plotted for three different decades of PC ranks using kernel density estimation. While there is a general trend for lower-rank PCs to have higher universality and brain similarity scores, there is nonetheless wide variation within each decade of PC ranks, with the scores from the first decade of ranks ranging as low as those from the third decade. 
    }
    \label{sup_14}
\end{figure}
\clearpage

\begin{figure}[H]
    \center
    \includegraphics[width=0.5\textwidth]{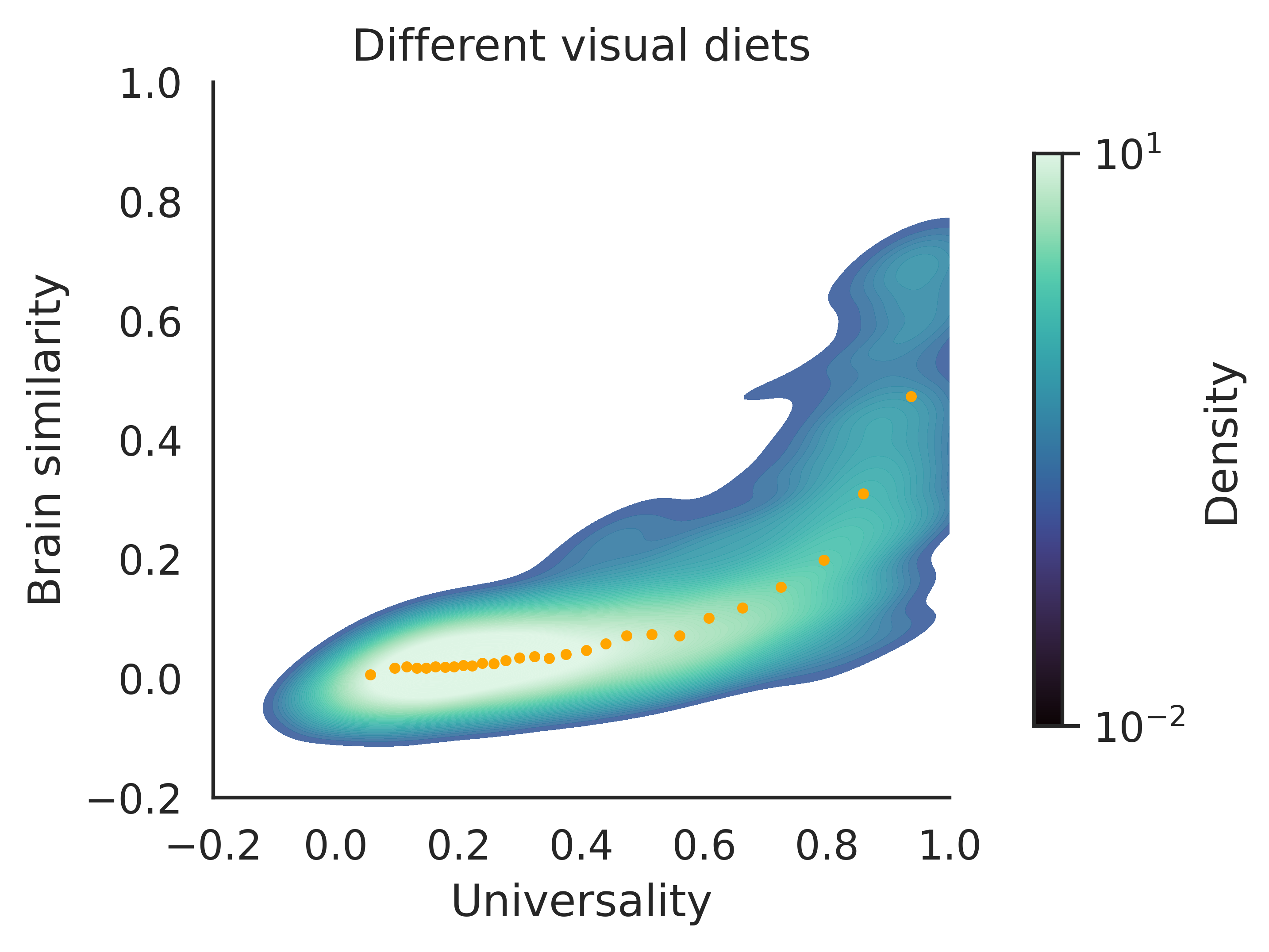}
    \caption{
    \textbf{Universality and brain similarity of neural networks trained with varied visual diets}
    This plot shows the relationship between universality and brain similarity of representational dimensions extracted from a set of models trained with varied visual diets but matched on architecture and task objective. The analysis is the same as in Figure \ref{fig_2}, and the plot shows the density of dimensions on a logarithmic scale computed using kernel density estimation. The orange dots show the mean universality and brain similarity scores for equally sized quantiles of 100 dimensions along the x-axis. As in Figure \ref{fig_2}, this plot exhibits a high density of points near the origin, showing that most dimensions are idiosyncratic to each network and are not shared with the human brain. However, there is also a subset of dimensions with exceptionally high universality and brain similarity scores. These latter dimensions correspond to representations that are consistently learned by networks with varied visual diets and that are also strongly shared with the visual representations of the human brain. 
    }
    \label{sup_13}
\end{figure}
\clearpage

\begin{table}[h]
    \caption{Networks from the set of trained models with varied architectures.}
    \centering
    \begin{tabular}{>{}c>{}c>{}c>{}c>{}c}
        \toprule
        Architecture             & Learning objective    & Architecture type & Training data & Source  \\
        \midrule
        ResNet18                 & Object classification & Convolutional     & ImageNet      & PyTorch \\
        ResNet50                 & Object classification & Convolutional     & ImageNet      & PyTorch \\
        ResNeXT50\_32x4d         & Object classification & Convolutional     & ImageNet      & PyTorch \\
        Wide\_ResNet50\_2        & Object classification & Convolutional     & ImageNet      & PyTorch \\
        AlexNet                  & Object classification & Convolutional     & ImageNet      & PyTorch \\
        VGG16                    & Object classification & Convolutional     & ImageNet      & PyTorch \\
        DenseNet121              & Object classification & Convolutional     & ImageNet      & PyTorch \\
        SqueezeNet1\_1           & Object classification & Convolutional     & ImageNet      & PyTorch \\
        ShuffleNet\_v2\_x1\_0    & Object classification & Convolutional     & ImageNet      & PyTorch \\
        ConveNeXt\_tiny          & Object classification & Convolutional     & ImageNet      & PyTorch \\
        Swin\_t                  & Object classification & Transformer       & ImageNet      & PyTorch \\
        MaxVit\_t                & Object classification & Transformer       & ImageNet      & PyTorch \\
        Cait\_xxs24\_224         & Object classification & Transformer       & ImageNet      & Timm    \\
        Coat\_lite\_tiny         & Object classification & Transformer       & ImageNet      & Timm    \\
        Deit\_tiny\_patch16\_224 & Object classification & Transformer       & ImageNet      & Timm    \\
        Levit\_128               & Object classification & Transformer       & ImageNet      & Timm    \\
        Mixer\_b16\_224          & Object classification & MLP-Mixer         & ImageNet      & Timm    \\
        ResMLP\_12\_224          & Object classification & MLP-Mixer         & ImageNet      & Timm    \\
        Dla34                    & Object classification & Convolutional     & ImageNet      & Timm    \\
        \bottomrule
    \end{tabular}
    \label{tab_2}
\end{table}

\begin{table}[h]
\caption{Networks from the set of trained models with varied training objectives.}
    \centering
    \begin{tabular}{>{}c>{}c>{}c>{}c>{}c}
        \toprule
        Architecture & Learning objective    & Training setting & Training data & Source  \\
        \midrule
        ResNet50     & Object classification & Supervised       & ImageNet      & PyTorch \\
        ResNet50     & Jigsaw                & Self-supervised  & ImageNet      & VISSL   \\
        ResNet50     & RotNet                & Self-supervised  & ImageNet      & VISSL   \\
        ResNet50     & ClusterFit            & Self-supervised  & ImageNet      & VISSL   \\
        ResNet50     & NPID++                & Self-supervised  & ImageNet      & VISSL   \\
        ResNet50     & PIRL                  & Self-supervised  & ImageNet      & VISSL   \\
        ResNet50     & SimCLR                & Self-supervised  & ImageNet      & VISSL   \\
        ResNet50     & SwAV                  & Self-supervised  & ImageNet      & VISSL   \\
        ResNet50     & DeepClusterV2         & Self-supervised  & ImageNet      & VISSL   \\
    \bottomrule
    \end{tabular}
    \label{tab_3}
\end{table}

\begin{table}[h]
\caption{Networks from the set of trained models with varied visual diets.}
    \centering
    \begin{tabular}{>{}c>{}c>{}c>{}c}
        \toprule
        Architecture & Learning objective  & Training data & Source  \\
        \midrule
        AlexNet-GN     & IPCL   & ImageNet      & IPCL \\
        AlexNet-GN     & IPCL   & OpenImages   & IPCL   \\
        AlexNet-GN     & IPCL   & Places2    & IPCL   \\
        AlexNet-GN     & IPCL   & VGGFace2     & IPCL   \\
    \bottomrule
    \end{tabular}
    \label{tab_4}
\end{table}

\end{document}